\documentclass[]{agujournal2019}
\usepackage{url} 
\usepackage{lineno}
\usepackage{soul}

\journalname{JGR: Planets}

\begin{document}

%
%

\title{Jupiter's Equatorial Plumes and Hot Spots:  Spectral Mapping from Gemini/TEXES and Juno/MWR}

%
%




\authors{L.N. Fletcher\affil{1}, G.S. Orton\affil{2}, T.K. Greathouse\affil{3}, J.H. Rogers\affil{4}, Z. Zhang\affil{2}, F.A. Oyafuso\affil{2}, G. Eichst\"{a}dt\affil{5}, H. Melin\affil{1}, C. Li\affil{6}, S.M. Levin\affil{2}, S. Bolton\affil{3}, M. Janssen\affil{2}, H-J. Mettig\affil{4}, D. Grassi\affil{7}, A. Mura\affil{7}, A. Adriani\affil{7}}

\affiliation{1}{School of Physics and Astronomy, University of Leicester, University Road, Leicester, LE1 7RH, UK.}
\affiliation{2}{Jet Propulsion Laboratory, California Institute of Technology, 4800 Oak Grove Drive, Pasadena, CA 91109, USA.}
\affiliation{3}{Southwest Research Institute, San Antonio, Texas, TX, USA.}
\affiliation{4}{JUPOS Team and British Astronomical Association, Burlington House, Piccadilly, London W1J ODU, UK}
\affiliation{5}{Independent Scholar, Stuttgart, Germany}
\affiliation{6}{Department of Astronomy, University of California Berkeley, Berkeley, CA 94720-3411, USA.}
\affiliation{7}{Istituto di Astrofisica e Planetologia Spaziali, Istituto Nazionale di Astrofisica, Roma, Italy}






\correspondingauthor{Leigh N. Fletcher}{leigh.fletcher@le.ac.uk}




\begin{keypoints}
\item Gemini TEXES spectral mapping reveals temperature, aerosol, and ammonia contrasts associated with plumes and hot spots on Jupiter's NEB jetstream.
\item Juno microwave measurements are consistent with the infrared mapping, and reveals that hot spot ammonia contrasts are confined to pressures less than 8-10 bars.
\item Hot spots and plumes are primarily contrasts in aerosols, with only subtle upper-tropospheric ammonia and temperature variations.
\end{keypoints}

%
%

%
%


\begin{abstract}

We present multi-wavelength measurements of the thermal, chemical, and cloud contrasts associated with the visibly dark formations (also known as 5-$\mu$m hot spots) and intervening bright plumes on the boundary between Jupiter's Equatorial Zone (EZ) and North Equatorial Belt (NEB). Observations made by the TEXES 5-20 $\mu$m spectrometer at the Gemini North Telescope in March 2017 reveal the upper-tropospheric properties of 12 hot spots, which are directly compared to measurements by Juno using the Microwave Radiometer (MWR), JIRAM at 5 $\mu$m, and JunoCam visible images.  MWR and thermal-infrared spectroscopic results are consistent near 0.7 bar.  Mid-infrared-derived aerosol opacity is consistent with that inferred from visible-albedo and 5-$\mu$m opacity maps.  Aerosol contrasts, the defining characteristics of the cloudy plumes and aerosol-depleted hot spots, are not a good proxy for microwave brightness.  The hot spots are neither uniformly warmer nor ammonia-depleted compared to their surroundings at $p<1$ bar.  At 0.7 bar, the microwave brightness at the edges of hot spots is comparable to other features within the NEB.  Conversely, hot spots are brighter at 1.5 bar, signifying either warm temperatures and/or depleted NH$_3$ at depth.  Temperatures and ammonia are spatially variable within the hot spots, so the precise location of the observations matters to their interpretation.  Reflective plumes sometimes have enhanced NH$_3$, cold temperatures, and elevated aerosol opacity, but each plume appears different. Neither plumes nor hot spots had microwave signatures in channels sensing $p>10$ bars, suggesting that the hot-spot/plume wave is a relatively shallow feature.

\end{abstract}

\section*{Plain Language Summary}
To date, our only direct measurement of Jupiter's gaseous composition came from the descent of the Galileo probe in 1995.  However, the results from Galileo appeared to be biased due to the unusual meteorological conditions of its entry location: a dark, cloud-free region just north of the equator, known as a hot spot.  One of the aims of NASA's Juno mission was to place the findings of the Galileo probe into broader context, which requires a detailed characterisation of these equatorial hot spots and their neighbouring plumes.  We combine (a) data from Juno (microwave observations sounding conditions below the clouds, and visible/infrared observations revealing variations in cloud opacity) with (b) observations from amateur observers (to track the hot spots over time) and (c) observations from the TEXES infrared spectrometer mounted on the Gemini-North telescope.  The latter provides the highest-resolution thermal maps of Jupiter's tropics ever obtained, and reveals contrasts within and between the individual hot spots and plumes.  We find that the hot spots are distinguishable from their surroundings for relatively shallow pressures, but that the deep measurements from Juno and Galileo are probably more representative of Jupiter's North Equatorial Belt than previously thought.

%
%

\section{Introduction}
\label{intro}

Jupiter's tropical domain is characterised by two eastward jet streams: the jet at $6.0^\circ$N (planetocentric latitude) that separates the red-brown North Equatorial Belt (NEB, $6.0-15.1^\circ$N) from the visibly-white Equatorial Zone (EZ, $6.2^\circ$S-$6.0^\circ$N); and the jet at $6.2^\circ$S that separates the South Equatorial Belt (SEB, $6.2-17.4^\circ$S) from the EZ \cite<see review by>{18sanchez_jets}.  These equatorial belts and zones exhibit remarkably different environmental conditions in the upper troposphere: the EZ is cold, typically cloud-covered, and exhibits enhancements in ammonia, phosphine, and other disequilibrium tracers such as para-H$_2$, whereas the NEB and SEB are warmer, with lower cloud opacities, and evidence for gaseous depletion \cite<see review by>{20fletcher_beltzone}.  NASA's Juno spacecraft \cite{17bolton} and ground-based millimetre/centimetre-wave observations \cite{16depater, 19depater_vla} have revealed that the belt/zone contrast in ammonia extends to great depths.  In particular, Juno's first close flyby (perijove 1, PJ1, on 27 August 2016) revealed that a column of enriched ammonia exists below the equatorial clouds \cite{17li}, consistent with the enriched ammonia observed in the upper troposphere \cite{06achterberg, 16fletcher_texes}.  However, the Juno-measured equatorial NH$_3$ abundance was at the lower end (but still within the uncertainties) of that derived from the Galileo probe during its descent to 22 bars in 1995 \cite{04wong_gal}, which was itself expected to be depleted compared to Jupiter's bulk abundance due to unique meteorological conditions at the entry site \cite{98orton}.  This begs the question of how representative the Juno and Galileo measurements are of Jupiter's tropics, and whether longitudinal contrasts \cite<or indeed temporal variability,>{18antunano_EZ} might be playing a key role.

The region surrounding the NEBs jet at $6.0^\circ$N, both in the northern EZ and the southern NEB, is one of the most longitudinally variable regions on the planet, owing to the existence of an equatorially-trapped Rossby wave on the NEBs jet \cite{90allison, 00showman, 05friedson}.  This has been thoroughly characterised in visible light, where a chain of $\sim10-13$ compact \cite<$3000\times10000$ km, >{13choi}, quasi-rectangular, and visibly-dark formations (DFs) spread around the full longitude circle of the NEBs \cite{98vasavada, 06arregi, 13choi, 19rogers}.  These DFs cover only 0.1-0.5\% of Jupiter's total area \cite{98ortiz}, but commonly, as in 2017, around 20-30\% of the longitude circle near $7^\circ$N.  The DFs are regions where low tropospheric cloud opacity \cite{98banfield} permit 5-$\mu$m radiance to escape from the 2-5 bar pressure levels, rendering them as bright `hot spots' in the infrared \cite{77terrile, 98ortiz}.  The DFs persist for many months, but can merge, split, and otherwise evolve with time as they move eastward along the NEBs jet at $\sim103$ m/s \cite{13choi}.  These features are thought to be associated with high-level convergence and subsidence, the dry downdrafts maintaining conditions that are depleted in clouds and volatiles \cite{98showman}.  

This pattern is suggestive of a planetary-scale wave with DFs at its troughs \cite{90allison, 98ortiz, 00showman, 05friedson}.  In between the DFs, at the crests of the planetary wave, the equatorial clouds are organised into white and reflective `fans' or `plumes' in the $2-6^\circ$N region \cite{76reese}, extending northeast from the equator to the NEBs where they appear to spread longitudinally, sometimes filling the longitudinal gap between DFs.  The brightest clouds are often seen at the northern edge of a plume, but not all plumes are the same, with some being brighter and `fresher' than others \cite{95rogers}.  The plume latitude is co-located with frequent detections of NH$_3$ ice \cite{02baines} and H$_2$O ice signatures \cite{00simon}, consistent with the idea of uplift.  The plumes are bordered to the southeast by darker `festoons,' which seem to emanate from the southwestern corner of the DFs and stretch southwest, and which become more vivid and easier to see during periods of EZ disturbances \cite<cloud-clearing events that occur once every 6-7 years,>{18antunano_EZ}.  East of the plume, and sometimes immediately south of a DF, anticyclonic gyres can be seen in the equatorial clouds, another potential manifestation of a Rossby wave \cite{05friedson} that may help to shape the morphology of the plumes and hot spots \cite{13choi}.  

If this equatorial wave governs the distributions of temperatures, clouds, and gaseous distributions, then both the Juno and Galileo measurements would depend upon which portion of the wave (plume, DF, or in between) that it sampled.  Indeed, the unexpected results from the Galileo probe are often ascribed to it entering the southern edge of a hot spot \cite{98orton}.  The fast eastward motion of the DFs and the short-term variation in their shapes, extents, and drift rates, makes it challenging for Juno to target a specific location in the wave, so the type of feature at the sub-spacecraft location must be determined a posteriori.  Confounding matters is the narrow longitudinal swath observed by Juno (around $2^\circ$ longitude at the equator during the first year of the mission), and the large time separation (53 days) between adjacent Juno measurements.  This study attempts to place Juno's microwave observations at tropical latitudes into context, by tracking plumes and hot spots at high spatial resolution.  Thermal infrared observations prior to Juno's arrival revealed that the plumes and dark formations influence the distribution of temperatures, aerosols, and ammonia in the troposphere above the clouds ($0.4<p<1.0$ bar), but had limited impact on the radiatively-controlled upper troposphere \cite<$p<0.4$ bar,>{16fletcher_texes}.  However, the spatial resolution of these Cassini and ground-based observations was limited, preventing direct comparison to high-resolution Juno observations.  We therefore performed thermal-infrared spectroscopy from the Gemini-North observatory in 2017, providing high-resolution thermal maps for direct comparison with Juno's 2017 observations.   

This article is organised as follows.  The sources of Juno and ground-based data are described in Section \ref{data}.  In Section \ref{amateur} we use a record of DF locations provided by amateur observers to predict whether or not Juno's perijove locations would come close to the desired features.  We then compare the amateur images to nadir-equivalent microwave brightness temperature maps derived from Juno's first eight perijoves to show that contrasts should exist from PJ to PJ in the EZ and NEB.  Given that Juno's microwave radiometer (MWR) only samples a narrow longitudinal swath, we also compare to JIRAM 5-$\mu$m observations and JunoCam visible-light observations.  The powerful combination of spectral mapping from TEXES with the diffraction-limited spatial resolution of Gemini's 8-m primary mirror enables mapping of temperatures, clouds and composition within the plumes and DFs for altitudes above the $\sim700-$mbar cloud deck in Sections \ref{texes} and \ref{maps}.  The TEXES results are used to predict the brightnesses in Juno's microwave observations in Section \ref{mwrmodel}.   Section \ref{discuss} shows that these results reveal internal contrasts within and between the DFs and plumes, and that (as of October 2017) Juno had yet to encounter a mature hot spot as depleted as that encountered by the Galileo Probe in 1995.    

\section{Data}
\label{data}

\subsection{Juno Observations}

This study employs three sources of data from the Juno spacecraft to investigate Jupiter's tropical plumes and dark formations (DFs):  observations from the Microwave Radiometer \cite<MWR,>{17janssen}, JunoCam \cite{17hansen}, and the JIRAM near-infrared instrument \cite{17adriani}.  These will be compared to amateur visible-light observations of Jupiter (see Section \ref{amateur}), a record of hot spot locations at 5 $\mu$m from NASA's Infrared Telescope Facility (IRTF, Section \ref{irtf_tracking}), and ground-based thermal-infrared spectral maps of Jupiter's tropics (see Section \ref{texes_data}).  MWR observations in six channels from 1.37 to 50 cm (0.6-22 GHz) are acquired as the spacecraft spins at 2 rpm during its $\sim2-$hour transit from the north to south pole \cite{17janssen}.  This means that the six antennae capture a range of emission angles for each latitude, such that the limb-darkening can provide a key constraint on Jupiter's 0.7-to-300-bar ammonia and water abundance.  However, the close proximity of Juno to Jupiter means that the longitudinal coverage is narrow, particularly at the equator, which is why the other data sources are used to provide spatial context.

\begin{figure}
\begin{centering}
\includegraphics[width=0.7\textwidth]{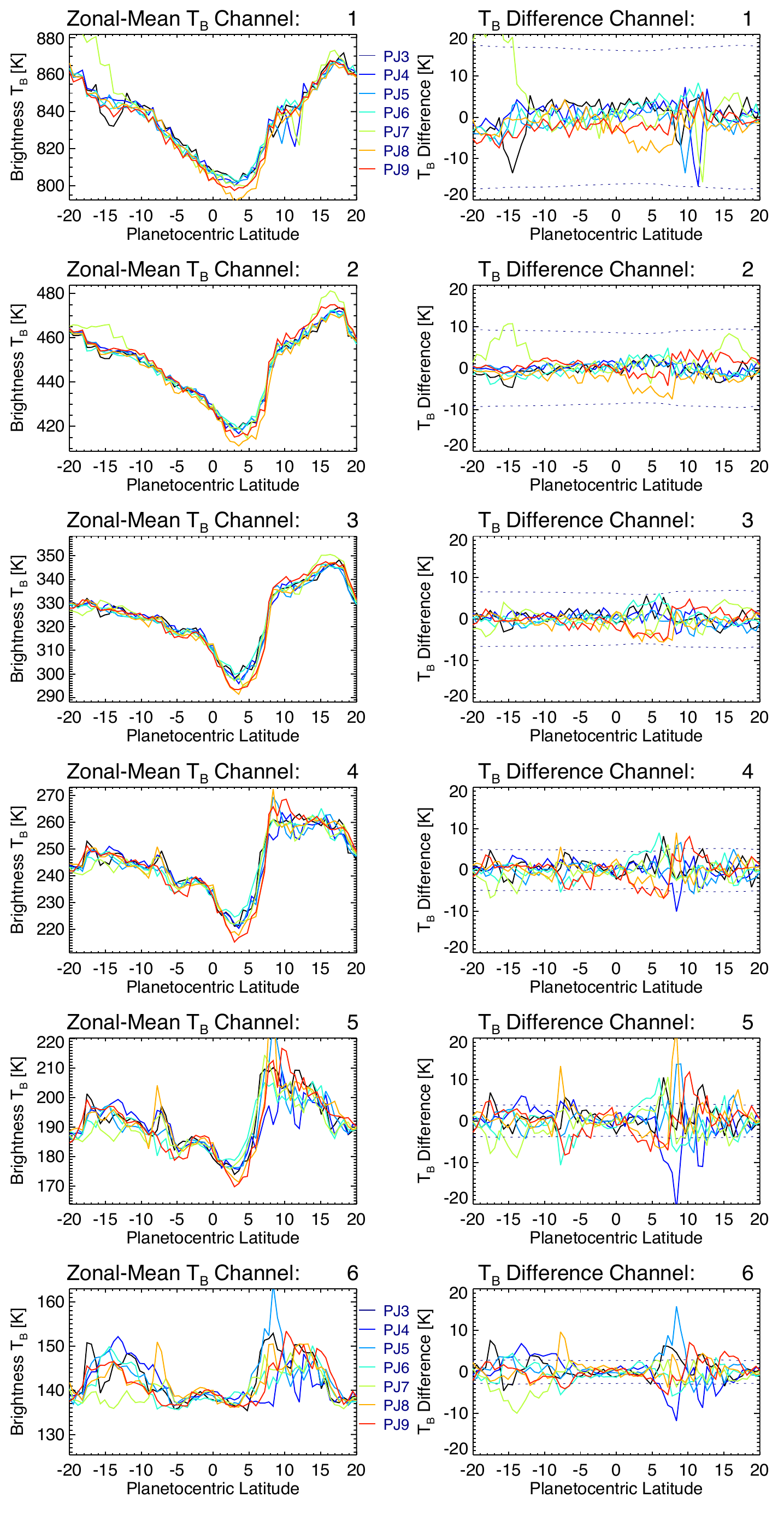}
\caption{Zonal-mean brightness temperatures in Jupiter's tropics measured by Juno MWR between PJ3 (December 2016) and PJ9 (October 2017) in each of the six radiometers, from Channel 1 (50 cm) to Channel 6 (1.37 cm).  The right-hand column shows the temperature difference between the individual perijove measurements and the mean of the PJ3-9 measurements.  The horizontal dotted lines in the right-hand column are conservative 2\% systematic uncertainties on the mean PJ3-9 brightness \cite{17janssen}.  However, instrumental contributions to variability are expected to be an order-of-magnitude smaller (described in the main text).} 
\label{mwr_zonal}
\end{centering}
\end{figure}

The MWR antenna temperatures measured by the six radiometers contain contributions from the planet in the main beam, the antenna side lobes, Jupiter's synchrotron radiation (mostly affecting the longest wavelengths), and the cosmic microwave background.  These contributions are deconvolved from the data using the algorithms described by \citeA{17janssen}, producing the brightness temperature at the boresight of the observation.  Limb darkening is represented via three coefficients fitted to the limb-darkening curve at each latitude, so that $T_B = \alpha_0 + \alpha_1(1-\mu) + \alpha_2(1-\mu)^2$, where $\mu=\cos\theta$ and $\theta$ is the emission angle.  The nadir brightness temperatures are represented by $\alpha_0$ and are shown in Fig. \ref{mwr_zonal}, where lower brightness temperatures can be interpreted as due to excess ammonia absorption or (at least in the short-wavelength channels) as reductions in kinetic temperature.  

Uncertainties in the coefficients depend on the spatial grid used for fitting the measured limb darkening, and a regularisation process is required as described by \citeA{19zhang}.  We initially follow \citeA{17li} in assigning a conservative 2\% uncertainty to the measured brightness temperatures in Fig. \ref{mwr_zonal}, representing the pre-launch absolute calibration testing of \citeA{17janssen}.  However, this is a systematic uncertainty in the MWR brightness that is constant with time during a PJ, so cannot account for any changes to the latitudinal dependence of the brightness.  Indeed, the consistency in Fig. \ref{mwr_zonal} from PJ to PJ at many latitudes testifies to the stability of both the atmosphere (at some latitudes) and the absolute calibration of MWR - we conservatively estimate that MWR instrumental effects contribute variability no larger than 0.2\%.  Variations in Fig. \ref{mwr_zonal} exceeding 0.2\% are therefore deemed to be a consequence of real spatial or temporal atmospheric variability.  

Although the zonally-averaged brightness in Fig. \ref{mwr_zonal} reveals nothing about the nature of the features within the boresight, they do reveal where Jupiter's atmosphere exhibits the most variability from PJ to PJ.  At 1.37 cm (sounding $\sim700$ mbar), Fig. \ref{mwr_zonal} shows that the lowest brightness is not located directly at the equator, but is offset to the  $2-5^\circ$N region \cite{17li}.  This is seen more clearly at higher pressures, and is a result of the column of enhanced ammonia that was first detected during PJ1 (27 August 2016) \cite{17li,17bolton}.  The 700-mbar region between $\pm4^\circ$ latitude appears to have been relatively stable over the period between December 2016 and October 2017 (differences $<2$ K), but this changes in the $5-10^\circ$N range, where there are considerable variations from the mean brightness, with a maximum difference of 25-30 K between the coolest and warmest brightness temperatures (PJ4 and PJ5, respectively) measured near $8^\circ$N.  This is the largest variability from perijove to perijove observed in the $\pm20^\circ$ latitude range of Jupiter's tropics, and larger than our conservative 0.2\% instrumental uncertainty envelope, suggesting that PJ4 (February 2017) might have sampled a cool, ammonia-rich plume.  The pattern is repeated at the 1.5-bar level sounded by Channel 5 (3.0 cm), where a $\sim40$-K contrast is observed between PJ4 and PJ8 at $8^\circ$N.  As we move to the 5-10 bar range (sounded by the 5.75 and 11.55-cm channels 4 and 3), the $5-10^\circ$N latitude range still stands out as a region of large variability, but the contrasts are more subdued, $\pm8$ K at 5.75 cm and $\pm5$ K at 11.55 cm.  This suggests that the contrasts associated with plumes and DFs becomes smaller with increasing pressure, being hard to distinguish for depths below $\sim10$ bar.

At the deepest pressures sounded by MWR at 24 and 50 cm (channels 2 and 1), the small-scale variability of the $p<10$ bar region is replaced by smoother latitudinal trends, with PJ8 and PJ9 being notably cooler than the previous measurements throughout the $2-8^\circ$N domain.  Given that this extends over a wider latitude range than the plumes/hot spots, we do not associate this trend with those dynamic features, and the analysis of this change will be part of a long-term assessment of MWR data.  Finally, our discussion so far has been restricted to the northern tropics, but some variability is observed in the SEB (albeit with lower contrast), and the PJ7 observations of the Great Red Spot are seen as cooler $T_B$ for latitudes poleward of $10^\circ$S (warmer $T_B$ for $p>10$ bar).


Zonally averaged brightnesses present a challenge when trying to understand what type of features were present in the main beam of each antenna.  In subsequent sections, we use the averaged limb-darkening coefficients from multiple perijoves to reconstruct nadir-equivalent brightness temperatures for longitudes within the MWR field of view \cite{19zhang}, with the caveat that separation of the limb darkening from true longitudinal variability is challenging.  Furthermore, we attempt to avoid synchrotron contributions to the maps by including only forward-look data for the southern hemisphere and after-look data for the northern hemisphere \cite{19zhang}.  We only consider MWR observations between PJ3 (December 2016) and PJ9 (October 2017) in this analysis, as listed in Table \ref{tab:pjs}.  These nadir-equivalent maps will be compared to visible-light imaging by the amateur community in Section \ref{amateur}.

\begin{table}
\caption{Juno perijoves considered in this study.  The System I and III west longitudes are provided for the equator-crossing time.  The difference between the two longitude systems changes by 7.364 deg/day (0.3068 deg/hour) \cite{95rogers}. }
\centering
\begin{tabular}{c c c c c}
\hline
PJ & Date & Equator Crossing UTC & Sys III Longitude & Sys I Longitude \\
\hline
3 & 11 Dec 2016 & 17:05 & 7.0 & 5.9 \\ 
4 & 02 Feb 2017 & 12:59 & 277.0 & 305.0 \\ 
5 & 27 Mar 2017 & 08:54 & 187.0 & 244.0 \\ 
6 & 19 May 2017 & 06:04 & 142.0 & 228.3 \\ 
7 & 11 Jul 2017 & 01:58 & 52.0 & 167.4 \\ 
8 & 01 Sep 2017 & 21:52 & 322.0 & 106.5 \\ 
9 & 24 Oct 2017 & 17:47 & 232.0 & 45.7 \\ 
\hline
\label{tab:pjs}
\end{tabular}
\end{table}

\subsection{Gemini TEXES}
\label{texes_data}
The TEXES instrument \cite<Texas Echelon Cross Echelle Spectrograph,>{02lacy} has proven to be a powerful means of characterising the atmospheric temperatures, composition, and aerosols on Jupiter during the Juno mission \cite{16fletcher_texes, 17cosentino, 18melin, 18blain}.  This cross-dispersed grating spectrograph provides spatially-resolved spectral maps in specially selected channels from the M band (5 $\mu$m), to the N band (7-13 $\mu$m), and the Q band (17-24 $\mu$m).  TEXES is typically mounted on NASA's Infrared Telescope Facility (IRTF), where the spatial resolution of the spectral cubes is limited by the diffraction pattern from the 3.1-m primary mirror.  These IRTF observations capture Jupiter's large-scale banded structure, vortices, and storms, but lack the resolution to explore the internal structure of the hot spots and plumes. In March 2017, TEXES was relocated to Gemini-North for an observing run capitalising on the improvement in diffraction-limited spatial resolution offered by the 8.2-m diameter of the primary mirror.  At this time, Jupiter was 4.6 AU from Earth, such that the diffraction-limited spatial resolution varied across the TEXES settings from 0.14" (470 km or 0.4$^\circ$ latitude at Jupiter's equator at 4.7 $\mu$m) to 0.57" (1880 km or $1.5^\circ$ latitude at Jupiter's equator at 18.6 $\mu$m), all sampled with a 0.5"-wide TEXES slit.  This high spatial resolution comes at the expense of diminished spatial coverage, and the 10-hour long programme focused exclusively on mapping Jupiter's tropics over $360^\circ$ of longitude.  These observations were made on March 12-14 2017, between Juno's PJ4 (February 2) and PJ5 (March 27), but they cover atmospheric features that are still recognisable in all of Juno's 2017 observations.

The 15-arcsec-long slit was aligned north-south, parallel to Jupiter's central meridian, and stepped across the planet from east to west.  A step size of 0.25 arcsec was used to Nyquist sample the 0.5-arcsec slit width.  Given the angular size of Jupiter in March 2017, this required approximately 160 steps, each with a 2-second integration time.  With 140 pixels along the TEXES slit, each scan therefore contains approximately 22,400 independent spectra.  Each scan was executed twice, consecutively, both to increase the signal-to-noise ratio and to allow the removal of any low-quality scan positions.  A block of nine spectral settings (listed in Section \ref{texes}) took approximately 75 minutes, so we repeated each setting three times per night during a 5-hour observing run (67,200 spectra per night), and then repeated again on the next night to cover $360^\circ$ of longitude (with more than a million independent spectra in each of the 9 settings).  Jupiter scan maps were acquired in seven groups (Table \ref{tab:data}) over three nights (12-14 March 2017).  Radiometric calibration was achieved using the difference between an internal flat field source and the sky emission, described in detail by \citeA{16fletcher_texes}, meaning that no standard stars were required as divisors.  Further details on the reduction process will be provided in Section \ref{texes}.

\begin{table}
\caption{Gemini/TEXES Observations.  The dates, times, and longitudes (System I and III west) are provided for each of the seven groups of observations.  Data for each group are available here: \url{https://doi.org/10.5281/zenodo.3702328}.}
\centering
\begin{tabular}{c c c c c}
\hline
Date & Group & Time Range (UT) & LCMIII & LCMI\\
\hline
2017-03-12 & 1 & 10:10-11:18 & 179.5-220.6 & 126.8-168.3\\
 & 2 & 11:26-12:30 & 225.5-264.2 & 173.2-212.2\\
 & 3 & 12:39-13:48 & 269.6-311.3 & 217.7-259.8\\
\hline
2017-03-13 & 4 & 10:14-11:15 & 332.6-009.5 & 287.3-324.5\\
 & 5 & 11:24-12:16 & 014.9-046.4 & 330.0-001.7\\
 & 6 & 12:35-13:36 & 057.9-094.7 & 013.3-050.5\\
\hline
2017-03-14 & 7 & 10:14-12:05 & 123.3-190.4 & 085.5-153.0\\
\hline
\label{tab:data}
\end{tabular}
\end{table}

\section{Identifying NEBs Features}
\label{tracking}

The chain of dark formations (DFs) and reflective plumes (PLs) moves eastwards (with respect to longitude System III) along the prograde $\sim114$ m/s NEBs jetstream at $6.0^\circ$N planetocentric latitude, with a velocity of approximately $\sim103$ m/s\footnote{This implies a westward-propagating wave in the frame of the prograde jet, approximated by the System-I longitude system described in the main text.}.  This means that a single feature can move some $\sim7^\circ$ of longitude in a 24-hour period, during which time it can also change shape and evolve.  Catching a DF or plume within Juno/MWR's sub-spacecraft field of view, which is limited to some $2^\circ$ in width near to the equator, requires a considerable amount of serendipity, and cannot be planned in advance.  Instead, we seek to reconstruct the atmospheric features beneath the perijove location using the record of features from the amateur observer community, as described in the following sections.  

\subsection{JUPOS Tracking}
\label{amateur}
\begin{figure}
\includegraphics{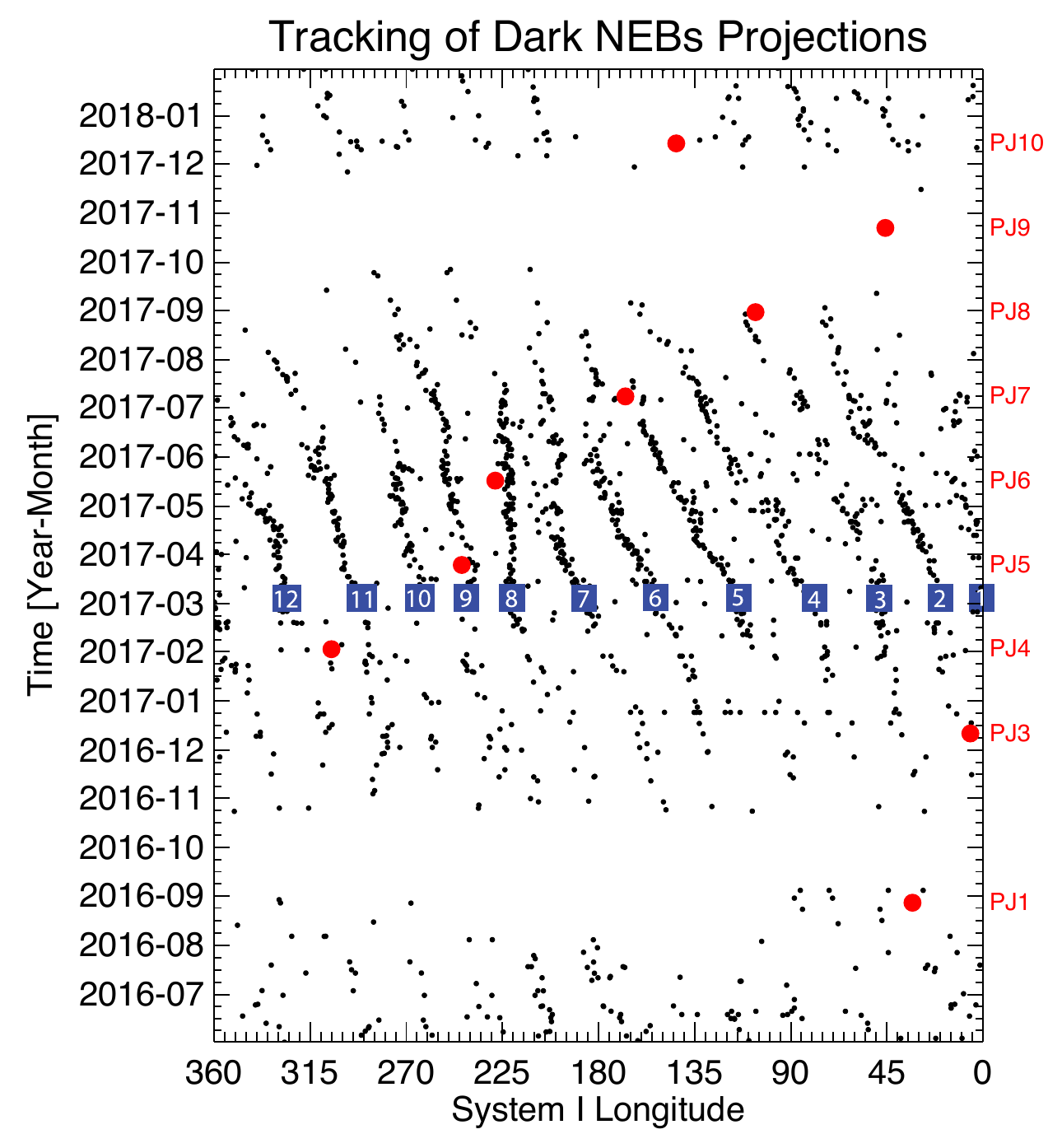}
\caption{Tracking the motions of `dark formations' on the NEBs using the JUPOS software.  Each black point is the longitude of a dark feature in amateur images of Jupiter, which are most frequent near opposition on 07 April 2017.  We use System-I longitude for ease of seeing the longevity of the features, and the individual dark formations are given an arbitrary number (referred to as `DFn' for the $n^{th}$ dark formation).  Reflective plumes (not shown) occur in between dark formations.  The System-I longitude of Juno's equatorial crossing are shown as red circles, and are labelled to the right.  No observations were acquired during PJ2, and only PJ3-9 are considered in this article. }
\label{jupos}
\end{figure}

During each apparition, a team of observers from around the world upload their near-nightly images of Jupiter to facilities such as the Planetary Virtual Observatory and Laboratory\footnote{\url{http://pvol2.ehu.eus/pvol2/}} \cite{18hueso} and the JunoCam website\footnote{\url{https://www.missionjuno.swri.edu/junocam/}} \cite{17hansen}.  The team behind the JUPOS project\footnote{\url{jupos.org}} use a freely-available software application to measure the position of discrete features on Jupiter via point-and-click, which are recorded in large databases subdivided by latitude range.  When plotted as a function of longitude and time, the locations trace out the lifetimes, motions, and evolution of these discrete features, albeit limited to (i) the resolutions afforded by amateur facilities \cite{14mousis_proam} and (ii) the ability of the users to identify faint details.  The rapid motion of the NEBs features across the disc still poses a challenge, so we transform from System III longitudes (i.e., the Voyager-defined jovian rotation rate) into System I longitudes.  The latter system, the first to be defined by jovian observers before the space age, is based on the observed motions of low-latitude features such as these, and the dark formations (DFs) are observed to drift slowly westward in this longitude system (Fig. \ref{jupos}).  

Robust detection of an NEBs dark formation can be seen where the clustered points in Fig. \ref{jupos} are at their most dense.  We label twelve individual dark formations that were identifiable in March 2017, at the time of the TEXES observations. Similarly, reflective white plumes exist in between the DFs, and we label them with the same number as the DF to their eastward side (e.g., PL5 is immediately west of DF5).  The red points in Fig. \ref{jupos} indicate the System-I longitude of Juno's equator crossing during each perijove (Table \ref{tab:pjs}), and allow us to make a first-order estimate of what the microwave radiometer should see.  PJ4 (February 2017) appears to be unique, far away from any hot spots (in between DF11 and DF12) and potentially coincident with plume PL11.  PJ5 and 6 (March and May 2017) occurred just to the west of hot spots DF9 and DF8, respectively, whereas PJ3 (December 2016, DF2), PJ7 (July 2017, DF5) and PJ8 (September 2017, DF3) all could have come close to hot spot features.  However, the resolution of these comparisons remains too coarse to be certain that Juno did encounter a hot spot or plume.   

\subsection{IRTF Tracking}
\label{irtf_tracking}

\begin{figure}
\includegraphics[angle=0,width=0.9\textwidth]{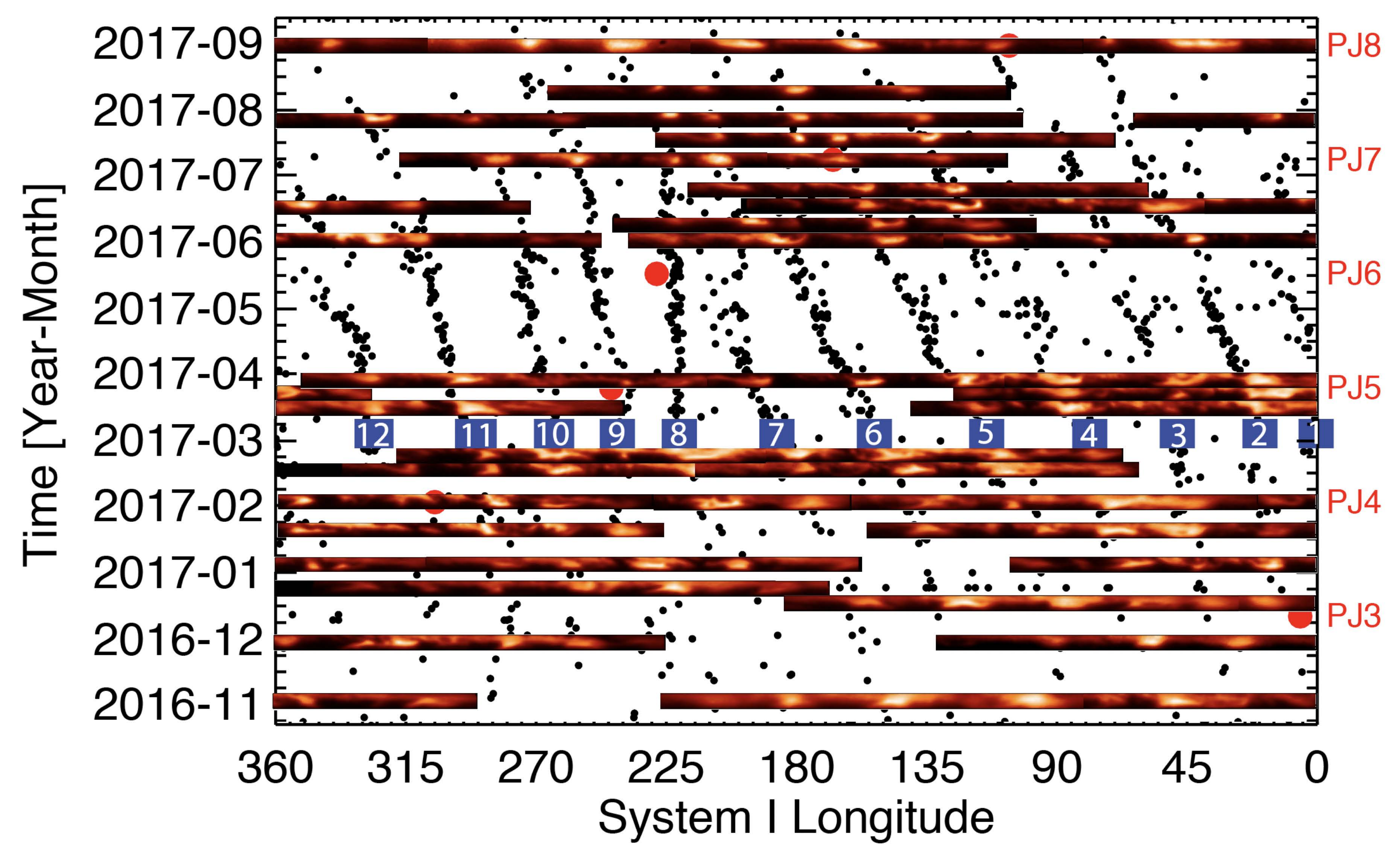}
\caption{IRTF/SpeX 5.1-$\mu$m observations of Jupiter's NEBs hot spots (DFs) in System-I longitudes, superimposed onto the amateur tracking in Fig. \ref{jupos}.  The images span $4-9^\circ$N planetocentric latitude, and observations over 2-3 days were combined to maximise longitudinal coverage. }
\label{irtf}
\end{figure}

Before proceeding with a comparison to the Juno observations, we first confirm that the JUPOS tracking of dark formations is consistent with the distribution and evolution of 5-$\mu$m hot spots.  Fig. \ref{irtf} superimposes 5.1-$\mu$m images of the NEBs hot spots acquired by the IRTF/SpeX instrument \cite{03rayner} onto the JUPOS tracking of the dark formations.  Processes for reducing and mapping the SpeX observations are described in \citeA{09fletcher_imaging}.  To the accuracy of the amateur and IRTF data, the cloud-free conditions responsible for the enhanced 5-$\mu$m brightness are co-located with the DFs in visible light.  Nevertheless, these images show how the morphology and contrasts of the DFs can change significantly over time, as found previously by \citeA{13choi}, meaning that comparisons to Juno observations should use images as close as possible in time.

\subsection{MWR Brightness Maps}
\label{mwrmaps}
\begin{figure*}
\includegraphics[angle=90,height=0.75\textheight]{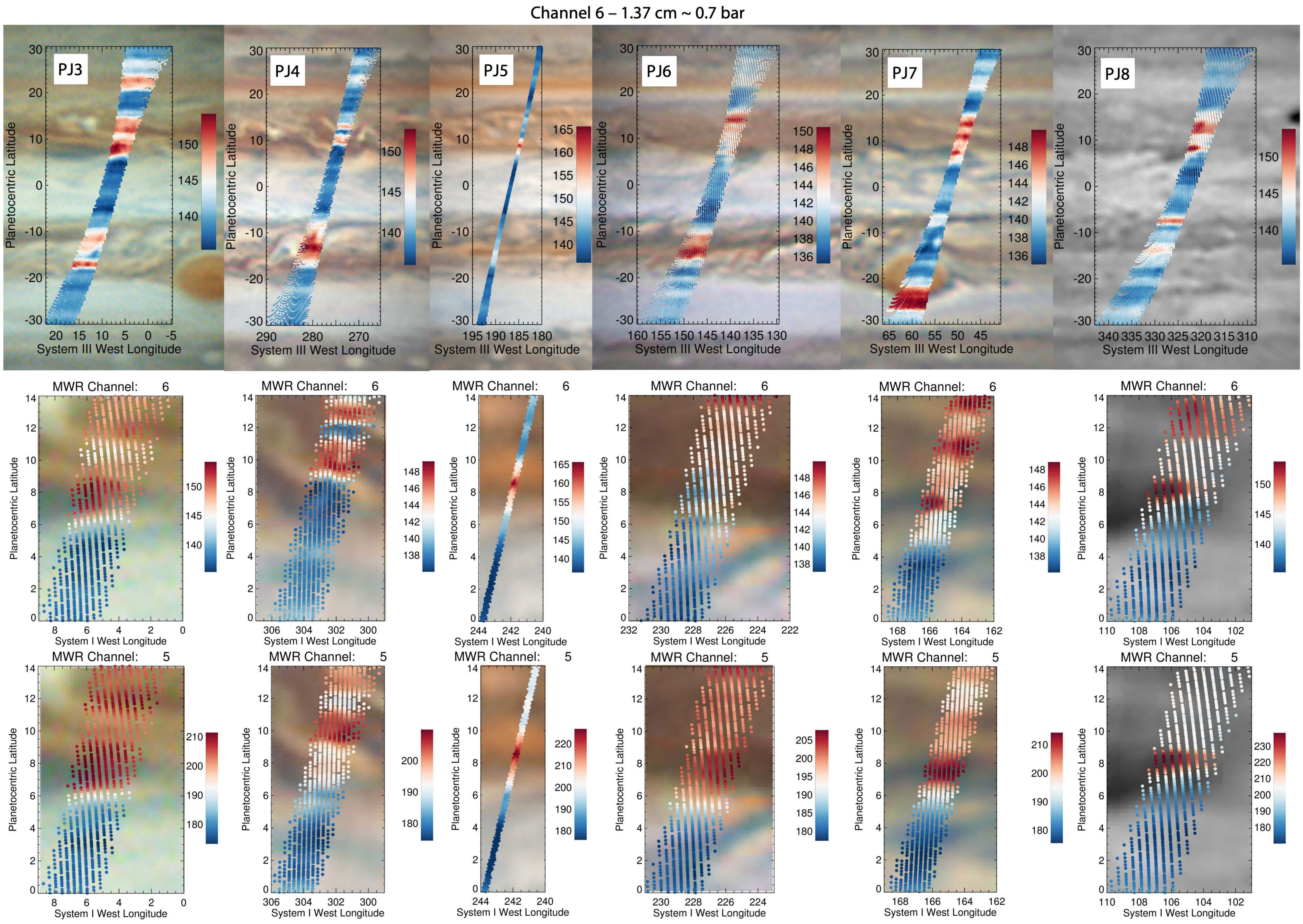}
\caption{Top Row:  Nadir-equivalent brightness temperatures in MWR channel 6 (1.37 cm) compared to amateur visible-light images acquired nearby in time. These images provide context for the MWR observations, and use System III longitudes.  However, time separations of a few hours can lead to substantial differences in location for the NEBs features.  Centre/Bottom Row:  To better compare the NEBs features, we transform both the amateur images and MWR observations to System-I longitudes.  Amateur images were acquired by the following observers:  I. Sharp 13 hours before PJ3 (11 December, 06:03UT); C. Foster 10 hours before PJ4 (2 February, 02:10UT); D. Peach within minutes of PJ5 (27 March, 08:52UT); C.Foster 12 hours before PJ6 (18 May, 18:31UT); C. Foster 11 hours before PJ7 (10 July, 15:11UT); and C. Foster 30 hours before PJ8 (31 August, 15:40UT).  Given the $30$-hour difference between the PJ8 observations and the amateur image, the MWR PJ8 map in the top row has been shifted by $9^\circ$ longitude to approximately account for the motion of the NEBs features.}
\label{mwr_img}
\end{figure*}

In Figure \ref{mwr_img}, the nadir-equivalent brightness temperature ($T_B$) maps from MWR channels 5 (3.0 cm) and 6 (1.4 cm), sensing 1.5 and 0.7 bar, respectively, are superimposed onto amateur observations of Jupiter acquired close in time.  We use System-III longitudes in the top row to provide context, and System 1 longitudes in the subsequent rows to precisely align features on the NEBs.  These maps confirm some of the conclusions from Section \ref{amateur}.  PJ3 (December 2016) shows (i) higher $T_B$ on the left side of the map, reaching 155 K at 1.4 cm at $8^\circ$N planetocentric at the eastern edge of a dark formation DF2 in the amateur image; and (ii) the coldest $T_B$ near $5^\circ$N, associated with a visibly-bright `gyre' of clouds in the amateur image.  This appears remarkably different from the next perijove, PJ4 (February 2017), where the coldest $T_B\sim136$ K span from $6-9^\circ$N and could be associated with an NH$_3$-rich and reflective equatorial plume, with no signs of any warm emission associated with a DF.  The NEB itself appears to be broken into warm and cool lanes (often referred to as `rifts'), with the coolest $T_B$ associated with visibly-bright clouds, and the warmest $T_B$ co-located with the deepest red colours\footnote{PJ4 is also notable for the MWR detection of longitudinal variability associated with a mid-SEB plume outbreak, as described by \citeA{19depater_alma}}.  The PJ5 $T_B$ map is much narrower than the previous two, spanning no more than a degree at the equator, and yet shows a significant warm region ($T_B\sim165$ K) near $8.5^\circ$N.  This is further north than the typical dark formations at $7^\circ$N, but appears to be associated with a dark horizontal band emanating from DF9 further to the east, near $175^\circ$W (System III), and we might suspect that brightness temperatures would have been even higher in the DF itself.

As inferred from the amateur tracking in Section \ref{amateur}, the track of PJ6 (May 2019) occurred just west of DF8.  Intriguingly, the dark features near $7-10^\circ$N do not show high contrast at 1.4 cm, but at 1.5 bar (channel 5) the $T_B$ reaches $\sim207$ K near the western edge of the DF8.  The track of PJ7 in July 2017 (notable for encountering the Great Red Spot) shows complex structure over the NEB, and appears to have passed directly over DF5.  At 1.4 cm, DF5 does not appear particularly bright when compared to other warm features within the NEB.  However, at 3.0 cm the dark material associated with the DF5 shows up as $T_B\sim213$ K, though still not as warm as the hot feature encountered during PJ5 at $8.5^\circ$N.  Finally, PJ8 (September 2017) occurred as Jupiter was nearing the end of the 2016/17 apparition, such that Earth-based imaging was considerably challenging.  Nevertheless, the IR image from Clyde Foster shows that PJ8 encountered the eastern edge of DF3 and saw the 3.0-cm brightness increase to $T_B\sim235$ K, the warmest encountered in this sequence.  Surprisingly, the 1.4-cm brightness does not exceed $T_B\sim154$ K, still cooler than the NEB hot feature encountered during PJ5. 

From the comparisons of MWR and amateur observations, we can draw the following conclusions:  (i) MWR may have encountered regions within DFs in PJ3, 6, 7 and 8, with the latter being the warmest in the 1.5-bar region; (ii) at 0.7 bar, the $T_B$ is often no warmer than the emission associated with other dark and red-brown features (e.g., NEB streaks), but has a higher contrast at 1.5 bar, and thus (iii) the DFs must have a complex vertical structure in the 0.7-1.5 bar range.  In the next section, we explore whether the DFs and bright MWR-emission are co-located with $5-\mu$m bright hot spots, and whether we can discern any structure at higher pressures.


\subsection{MWR-JIRAM Comparison and Hot Spot Depth}

\begin{figure*}
\includegraphics[angle=0,width=1.2\textwidth]{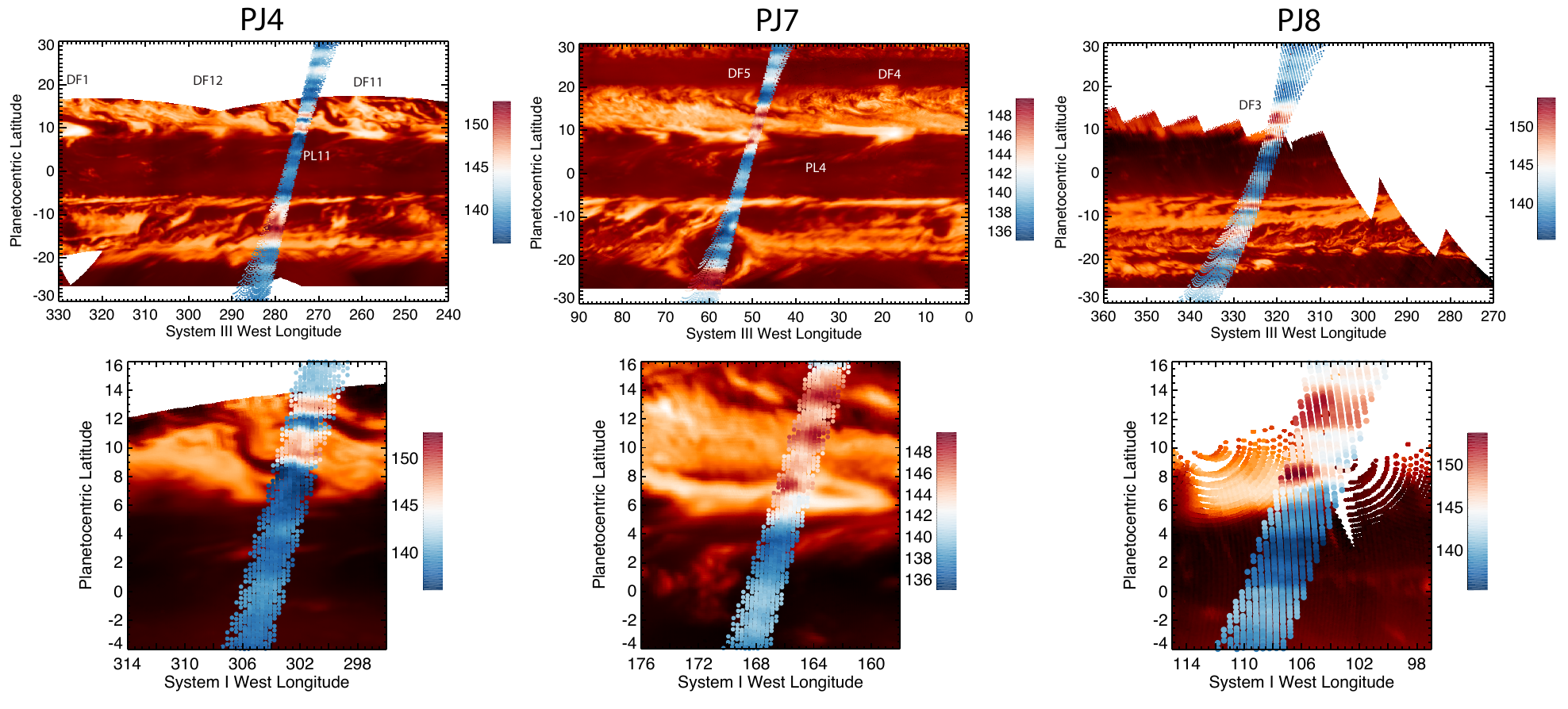}
\caption{M-band (5-$\mu$m) maps from Juno/JIRAM acquired in the hours before or after Juno perijoves for PJ4 (left, 02-Feb-2017, acquired 10-17 hours before PJ), PJ7 (centre, 11-Jul-2017, acquired 5-17 hours before PJ) and PJ8 (right, 01-Sep-2017, acquired 1-2 hours before PJ and 1-6 hours after PJ).  These are compared with MWR brightness temperatures at 1.4 cm acquired at each PJ (the legend refers to the temperatures in the MWR scan).  System-III maps are adequate for comparing MWR and JIRAM data away from the equator, but NEBs features move eastward by approximately $0.3^\circ$/hour, so a proper comparison requires transformation of the JIRAM maps into System 1, which is shown in the bottom row of this diagram.  The JIRAM maps have been stretched logarithmically to accentuate faint features in the EZ, but the 5-$\mu$m brightness temperatures range from 165-248 K.  JIRAM observations on PJ3, PJ5 and PJ6 did not cover the same area as MWR.}
\label{mwr_jiram}
\end{figure*}

The complexity of Juno's orbital track and spinning motion means that the JIRAM M-band (5 $\mu$m) imager does not always map the same region on Jupiter as the MWR instrument.  We surveyed JIRAM maps from PJ3 to PJ8, and found that only PJ4, PJ7 and (to a lesser extent) PJ8 covered the EZ/NEB region at the same longitudes as the MWR scans.  Fig. \ref{mwr_jiram} compares the 1.4-cm MWR brightness temperatures to M-band maps, showcasing the three best examples of the features of interest to this study - a cold plume PL11 in PJ4, the centre of DF5 in PJ7, and the eastern edge of DF3 in PJ8. We have applied a logarithmic stretch to the JIRAM data to accentuate fainter features in the EZ, and have employed both System-III and System-I longitudes to allow for intercomparisons (JIRAM data were typically taken a few hours ahead of the perijove).  In general, warm 1.4-cm emission coincides with regions that are bright (i.e., cloud-free) at 5 $\mu$m (and vice versa), although the structure is complex in the NEB.  This suggests that 5-$\mu$m brightness (related to aerosols) is not a good proxy for high microwave brightness temperatures (related to ammonia and temperature).  The PJ4 track encountered plume PL11 (to the west of DF11 in Fig. \ref{jupos}), which was dark and cloudy at 5-$\mu$m and cold (i.e., either ammonia-rich or physically cool) at 1.4 cm.  

Fig. \ref{mwr_jiram} shows that the PJ7 track did indeed encounter a 5-$\mu$m hot spot (DF5), but the brightest emission was confined to a small area near $7.5^\circ$N.  Unfortunately the M-band map during PJ8 only just encounters DF3, but confirms that the MWR scan did encounter the western edge of this feature.  We showed in Fig. \ref{mwr_img} that the hot spots exhibit more contrast in Channel 5 (3.0 cm, sounding 1.5 bar) than in Channel 6 (1.4 cm, sounding 0.7 bar).   However, the peak 5-$\mu$m brightness ($\sim250$ K) is warmer than all of the 3-cm observations sounding 1.5 bar (maximum of $T_B\sim235$ K during PJ8), and is closer to the 260-270 K range observed by MWR channel 4 (5.75 cm), sounding $\sim3.5$ bar \cite<note that cloud-free M-band contribution functions probe near or below the 4-bar level, Fig. 1 of>{18bjoraker}.  For PJ8, the hot spot was still visible in Channel 4 (5.75 cm) with $T_B=270$ K, but by Channel 3 (11.55 cm, sounding $\sim10$ bar) it could not be reliably distinguished from the rest of the NEB.  For PJ7, the hot spot lost contrast in Channel 4 (5.75 cm), and was indistinguishable from the surrounding NEB by Channel 3 (11.55 cm).

From this very small sample, we suggest that the hot spots are indistinguishable from the surrounding atmosphere at the depths probed at 11.55 cm (approximately 10 bar).  The ammonia contrasts within the hot spots and plumes are therefore thought to be `weather-layer' features restricted to $p<10$ bar, although MWR observations with wider longitudinal coverage would help to confirm this.  This is qualitatively consistent with the VLA-derived maps of Jupiter by \citeA{16depater} and \citeA{19depater_vla}, which showed evidence for hot spot/plume structures in the `radio-hot belt' at the NEBs, at least to depths of $\sim8$ bars sounded by their 4-8 GHz (3.7-7.5 cm) maps.  The shallow depth of the hot spots suggests that both the Galileo and Juno measurements of the regions deeper than 10 bar could be more representative of the wider NEB than previously thought (see Section \ref{discuss}).

\subsection{MWR-JunoCam Comparison}

\begin{figure*}
\includegraphics[angle=90,height=0.85\textheight]{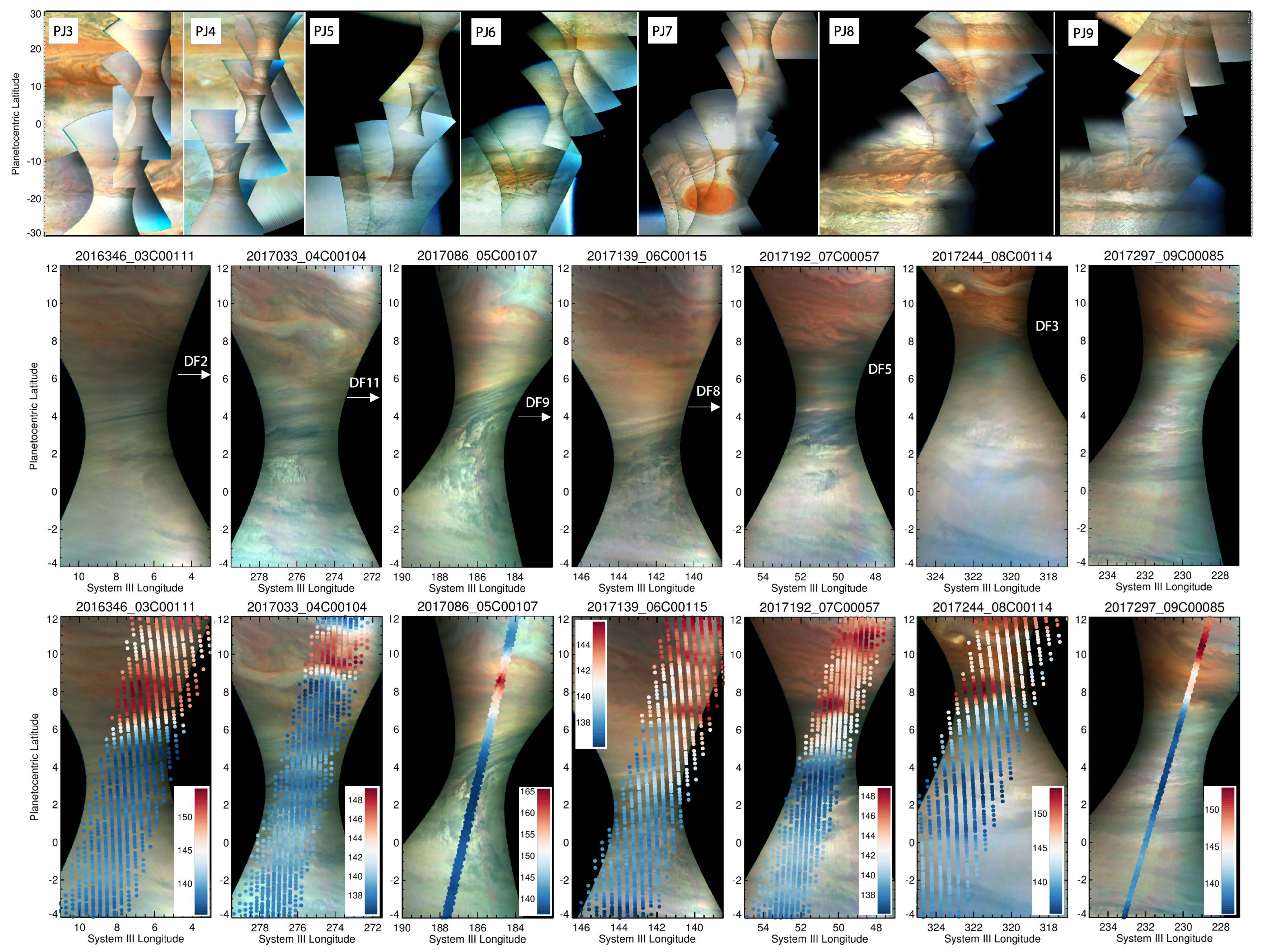}
\caption{JunoCam visible-light imaging of Jupiter's tropics during PJ3-9.  The top row features a montage of multiple JunoCam images, geometrically registered but then manually adjusted for optimum fit of cloud features, spanning $\pm30^\circ$ latitude.  The closest-approach image is then mapped individually in the centre row ($4^\circ$S to $12^\circ$N) to reveal the small-scale structure of the plumes, dark formations, and festoons.  The relevant DF is marked, based on Fig. \ref{jupos}.  The title for each panel is in the format YYYYDDD\_PJCFFFFF, where YYYY is the year, DDD is the day of the year, PJ is the perijove number, and FFFFF is the file number.  Finally, the bottom row replicates the central row, but overplots the MWR brightness at 1.4 cm.}
\label{junocam}
\end{figure*}

As a final check of the features within Juno's field of view, Fig. \ref{junocam} shows JunoCam visible-light images of the tropics taken near-simultaneously with the MWR scans.  JunoCam is a push-frame imager, using Juno's rotation to sweep the four filter strips (red, green, blue, and 890-nm) across the target \cite{17hansen}.  The $58^\circ$-wide field of view covers only a narrow longitude range near perijove, resulting in the `hour-glass' appearance of the mapped products in Fig. \ref{junocam}.  We employ an image-processing pipeline developed by G. Eichst\"{a}dt and described by \citeA{17orton} and \citeA{20tabatata}, whereby the orientation, optical properties, and observation timings are optimised via manual comparison of the observed limb positions.  However, there could be significant uncertainties in the geometric registration at low latitudes - the montage maps on the top row of Fig. \ref{junocam} required manual co-alignment of features observed in overlapping images.  Despite these uncertainties, the RGB images provide a useful guide for the features at Juno's sub-spacecraft point taken at the same time as the JIRAM and MWR observations.

The central row of Fig. \ref{junocam} reveals the NEBs features at high resolution from altitudes of 3400-4200 km above the 1-bar level.  The images have been divided by a fitted fourth-order polynomial to approximately adjust for strong illumination variations across the image, and the results should not be taken as true-colour.  In Section \ref{mwrmaps} we suggested that MWR may have encountered the edges of dark formations on PJ3, 7 and 8.  This is certainly true for PJ8, where the eastern edge of DF3 is visible, along with an expanse of bright clouds to the south that may be coincident with an anticyclonic gyre.  It is also true for PJ7, where a dark and complex region within DF5 can be seen at $7^\circ$N, but also the dark striations of a festoon between $2-4^\circ$N, with attendant small-scale white clouds.  But for PJ3, Fig. \ref{junocam} suggests that Juno observed blue-grey festoons emanating from DF2, rather than the centre of the hot spot itself, and indeed the whole region in Fig. \ref{mwr_img} appears warm.  Further festoons are seen for PJ6 (following DF8) and PJ5 (following DF9), the latter showing considerable structure within the blue-grey festoon streaks, along with small scale outbursts of clouds where the festoon meets the rest of the EZ.  These features are too small to resolve in the MWR maps. 

Section \ref{mwrmaps} and Fig. \ref{mwr_jiram} suggested that PJ4 flew over a plume, PL11.  This is somewhat unclear from Fig. \ref{junocam}, which shows a blend of blue-grey features and `flocculent' white clouds that even cast shadows, but these features must be sufficiently opaque to block the 5-$\mu$m brightness in Fig. \ref{mwr_jiram}.  Finally, the complexity of the NEB is shown in the $9-12^\circ$N range in Fig. \ref{junocam}, displaying white rifts that are stretched from northwest to southeast by the meridional shear on the zonal winds.  One such rift is particularly well defined in PJ8, along with white clouds in PJ4 and 5.  This high degree of variability, if mirrored in the temperature and ammonia distributions (see Section \ref{texes}), could be responsible for the contrasts in the brightness observed in MWR channels 5 and 6 in Fig. \ref{mwr_img}.

\section{Gemini/TEXES Observations}
\label{texes}

The limited longitudinal coverage afforded by the MWR observations presented a challenge when trying to compare plume and dark formation (DF) conditions to their surroundings in Section \ref{tracking}.  In this section, we employ mid-infrared spectroscopic mapping to characterise the temperatures, gaseous composition (ammonia, phosphine) and aerosol opacity in the upper troposphere ($p<0.8$ bar), in the altitude domain overlapping with the short-wave MWR Channel 6 (1.37 cm).  Although this lacks sensitivity to structures beneath the topmost condensation clouds, the ground-based maps have the benefit of full longitudinal coverage over a short period of time.  As described in Section \ref{texes_data}, the TEXES instrument was temporarily relocated to the Gemini-North observatory in March 2017, in between PJ4 and PJ5. Via efficient east-west scan-mapping, this generated some of the highest-spatial-resolution mid-infrared (5-20 $\mu$m) spectral maps of Jupiter ever obtained.  The maps spanned approximately $30^\circ$N to $30^\circ$S and were acquired in seven distinct groups (a single set of nine TEXES settings, Table \ref{tab:data}) between March 12 and March 14 2017, and in nine spectral settings similar to those listed in Table 1 of \citeA{16fletcher_texes}.  Contribution functions are shown in Fig. 7 of that paper, and reveal that the data sound:
\begin{itemize}
\item Tropospheric temperatures using the H$_2$-He collision-induced absorption at $537-539$, $585-589$ and $742-747$ cm$^{-1}$ (sensing the 100-500 mbar range); 
\item Stratospheric temperatures using methane emission from 1245 to 1250 cm$^{-1}$ (sensing 0.01-20 mbar); 
\item Tropospheric ammonia and phosphine from $893-911$ and $960-978$ cm$^{-1}$ (sensing 300-700 mbar); 
\item Upper-tropospheric aerosol from 1155 to 1188 cm$^{-1}$ (sensing 500-1000 mbar); 
\item Deep tropospheric aerosol from 2132 to 2142 cm$^{-1}$ (sensing $p>4$ bars in the aerosol-free case); and 
\item Stratospheric ethane (from 816 to 822 cm$^{-1}$) and acetylene (from 742 to 747 cm$^{-1}$) in the 0.01-10 mbar range.
\end{itemize}
Note that the eight spectral settings from 539 to 1248 cm$^{-1}$ are fitted simultaneously to derive temperature, composition, and aerosol distributions in the $p<1$ bar region in the following sections.

\subsection{Mapping of spectral cubes}

Assignment of latitudes and longitudes to each TEXES spectrum is rendered challenging by two factors - the exquisite spatial resolution, and the inability to see the planetary limb at each step in the scan.  Automated mapping software applied in previous IRTF observations was found to be accurate to $\sim3^\circ$ of longitude, but fine-tuning of consecutive scan maps was required to ensure that discrete atmospheric features lined up.  These adjustments were performed manually on a case-by-case basis - longitude shifts were determined for features identified in the South Equatorial Belt that were not expected to move during the interval spanned by this dataset.  These same shifts were applied to the northern hemisphere to check that they did not produce spurious results.  The shifts were estimated only for groups taken within an hour or two of each other in Table \ref{tab:data}.

Once the spectral cubes had been destriped (removal of short-term telluric variability in each scan, see \ref{destripe}), radiometrically scaled, and re-aligned, they were interpolated onto a regular grid for mapping.  For spectra that were greatly affected by telluric absorption, the difference in Doppler shift between the dawn and dusk limbs could be significant, with the consequence that some bright contributions to the spectral average might be invisible on one limb, but prominent on the other.  This sometimes produces east-west asymmetries in a single image cube that can only be removed if we discard all spectral regions affected by tellurics.  For this reason, we do not attempt to show complete $360^\circ$-longitude maps for all filters.  

\begin{figure*}
\includegraphics[angle=0,width=1.0\textwidth]{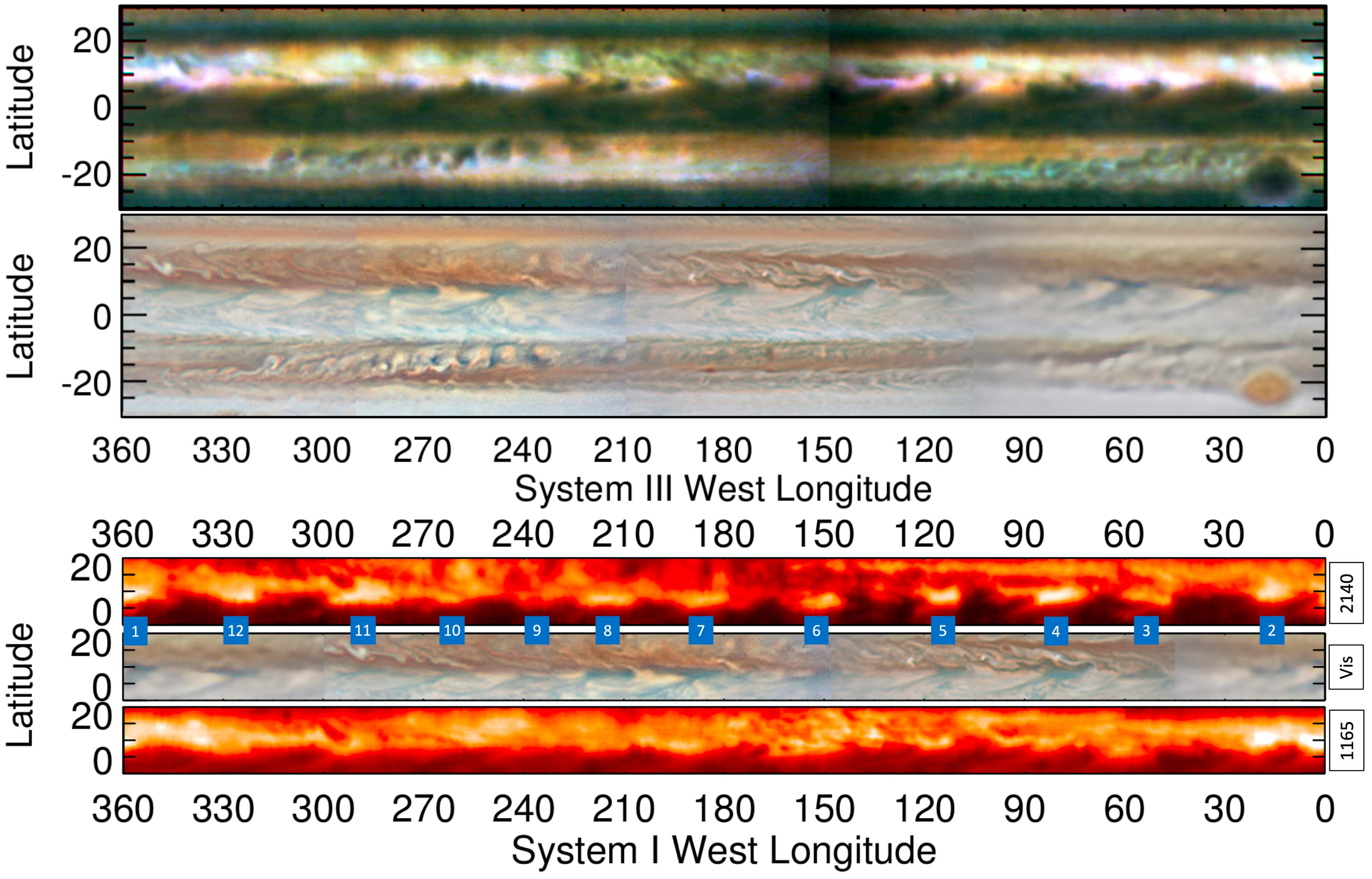}
\caption{\textbf{Top:}  Comparison of Gemini/TEXES 3-colour composite image (March 12-13) to amateur observations of Jupiter (March 10-11) in System III Longitude.  The TEXES image was constructed by summing radiances in three ranges:  $B=1158-1160$, $G=1160-1163$, and $R=1166-1168$ cm$^{-1}$. In this scheme, deep features dominate areas that are red and higher features dominate areas that are blue.  The vertical `seam' near $150^\circ$W is an artefact of the mapping process.  The amateur image was produced by M. Vedovato using observations from T. Olivetti, T. Kumamori, and E. Martinez\footnote{\url{http://pianeti.uai.it/index.php/Giove:_Mappe}}.  \textbf{Bottom:}  Given the fast System-III motion of NEBs features, we compare the visible map to radiances averaged over the full 1165 cm$^{-1}$ (8.6 $\mu$m) and 2140 cm$^{-1}$ (4.7 $\mu$m) settings in System I longitudes.  A logarithmic scale was used for the latter setting to make low-contrast features visible.  The dark formations are labelled using the numbering scheme from Fig. \ref{jupos}.  Latitudes are planetographic. }  
\label{map1165}
\end{figure*}

However, the 1165-cm$^{-1}$ setting provides the highest-resolution view of the tropospheric features with minimal telluric contamination, so we constructed a crude three-colour image in Fig. \ref{map1165} to demonstrate the powerful combination of Gemini and TEXES.  This is compared to visible light imaging from amateur observers acquired 24-48 hours earlier.  The red, green and blue channels were selected to probe from high pressures ($\sim0.6$ bar) to lower pressures ($\sim0.2$ bar).  Hot spots appear white in all channels, whereas plumes are dark in all channels, indicating structures that persist over the full altitude range.  Festoons emanating south-west from the hot spots appear brighter (i.e., thinner clouds) than the rest of the dark EZ.  The fact that we see colour in this figure at all shows how the spectrum changes from point to point:  for example, the dark and cloudy plumes in the SEB between 240 and $300^\circ$W associated with a mid-SEB outbreak \cite{17fletcher_seb, 19depater_alma}, and in the turbulent wake of the GRS between 30 and $90^\circ$W \cite{10fletcher_grs}, in contrast to spectra with thinner clouds appearing `red' in the rest of the SEB.  Furthermore, in the NEB there is a notable difference between 150 and $220^\circ$W, where prominent rifting activity generated thicker $\sim600$-mbar clouds (white in visible light) contrasted to the rest of the cloud-free NEB.  These colour contrasts within a single TEXES setting demonstrate the information content of these data, which will be harnessed via spectral inversions in Section \ref{maps}.

\begin{figure*}
\includegraphics[angle=0,width=1.0\textwidth]{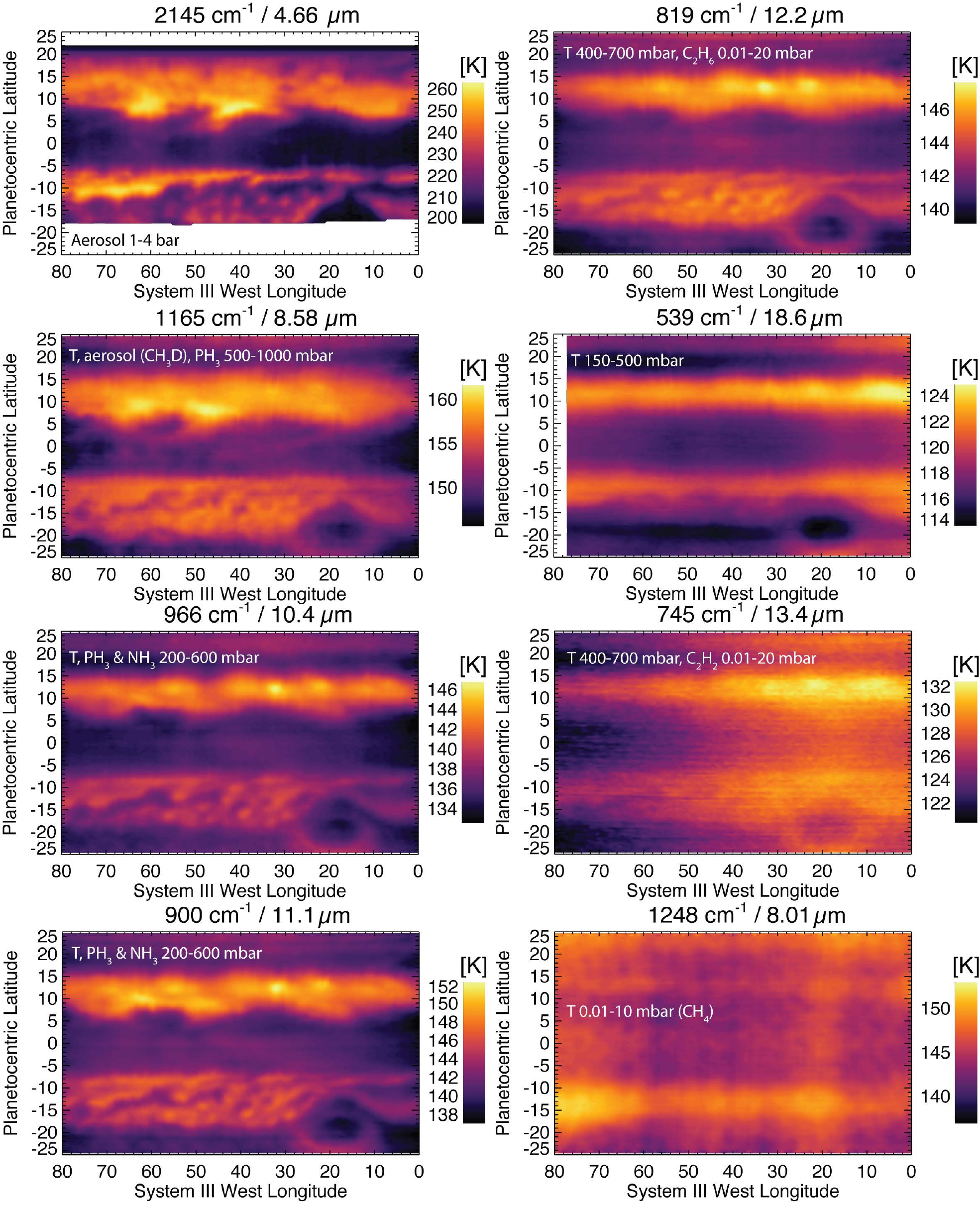}
\caption{An example brightness temperature maps of a single group (group 5) of TEXES scan maps, in eight of the nine channels (587 cm$^{-1}$ is omitted as it looks almost identical to 539 cm$^{-1}$).  This group features the GRS near $20^\circ$W and several plumes and hot spots on the NEBs near $6^\circ$N.  Some residual striping is evident in the 1248-cm$^{-1}$ setting, and the narrower slit at 2145 cm$^{-1}$ cuts off latitudes poleswards of $\pm18^\circ$. Labels on each panel reveal the key species being sensed in each setting.  Maps of all seven groups (without labels) in Table \ref{tab:data} are available in the Supplemental Material.}
\label{group5}
\end{figure*}

Figure \ref{group5} showcases maps of a single group (Group 5 in Table \ref{tab:data}) for eight of the nine spectral settings (maps for all seven groups are available in the Supplemental Material at \url{https://doi.org/10.5281/zenodo.3706796}).  The appearance of the plumes and hot spots on the NEBs is rather different from setting to setting, being undetectable at 539, 587, 745 and 1248 cm$^{-1}$, thus confirming that they are a feature of the atmosphere for $p>300$ mbar, with no signature at higher altitudes.  In some of the groups, particularly sensing the $240-330^\circ$W range in Fig. \ref{map1165}, a faint thermal wave can be observed over the mid-NEB.  This is at a higher latitude and distinct from the NEBs features, and has been present at a variety of epochs, often during a period of NEB expansion \cite<see>[for full details]{17fletcher_neb}.  

\subsection{Zonal mean inversions}

The TEXES maps will be inverted to map contrasts in temperatures, gaseous abundance, and aerosol opacity associated with the NEBs plumes and DFs for comparison to the MWR data.  This requires two precursors: (i) modifications to radiometric calibration for consistency with Cassini spectroscopy; and (ii) cross-checking of the zonally-averaged atmospheric properties with previous studies.  The modelling below uses the radiative transfer and optimal estimation retrieval software, NEMESIS \cite{08irwin}, which has been previously applied to TEXES mapping of Jupiter by \citeA{16fletcher_texes}.  The precise wavelength coverage of the TEXES channels differs between the IRTF and Gemini observations, so $k$-distributions (ranked lists of gaseous line data covering the temperature and pressure range relevant to Jupiter) were recomputed for each spectral setting, using the sources of spectroscopic line data in Table 2 of \citeA{16fletcher_texes}.   

The calibration cross-check requires the use of the best-fitting atmospheric profiles from Cassini Composite Infrared Spectrometer (CIRS) observations of Jupiter in December 2000 (see \ref{ref_model}).  These are used to forward-model the expected radiances in each TEXES channel for every latitude and observing geometry.  The TEXES observations are then scaled so that any `pseudo-continuum' in each channel is made to match the Cassini-derived forward model for latitudes $<\pm30^\circ$.  This typically requires 50-80\% reductions in the calibrated TEXES brightness \cite<e.g., see Fig. 8 of>{16fletcher_texes}, and is a known and repeatable feature of TEXES calibration in low- and medium-resolution settings, where the detectors exhibit non-linear behaviour as a function of brightness \cite{18melin}.  The largest changes were found in the middle of the N-band near 10 $\mu$m (1000 cm$^{-1}$), where the TEXES data had to be reduced by 7.8-9.3 K (in terms of brightness temperature).  Note that, following \citeA{18melin}, we do not adjust the measured radiances in the 1248 cm$^{-1}$ channel.  The CIRS reference model of \citeA{16fletcher_texes} has been updated in this work in an attempt to find consistency with the zonally-averaged NH$_3$ profiles derived from MWR data acquired during PJ1 (27 August 2016) by \citeA{17li} - full details are given in \ref{ref_model}. We find that the MWR-derived profiles contain too little NH$_3$ in the upper troposphere to explain the depth of the mid-IR absorption features (likely due to the choice of $T(p)$ in the early MWR analysis), and therefore we continue to use CIRS to guide our prior for the TEXES calibration and retrievals.

\begin{figure*}
\begin{center}
\includegraphics[angle=0,width=1.2\textwidth]{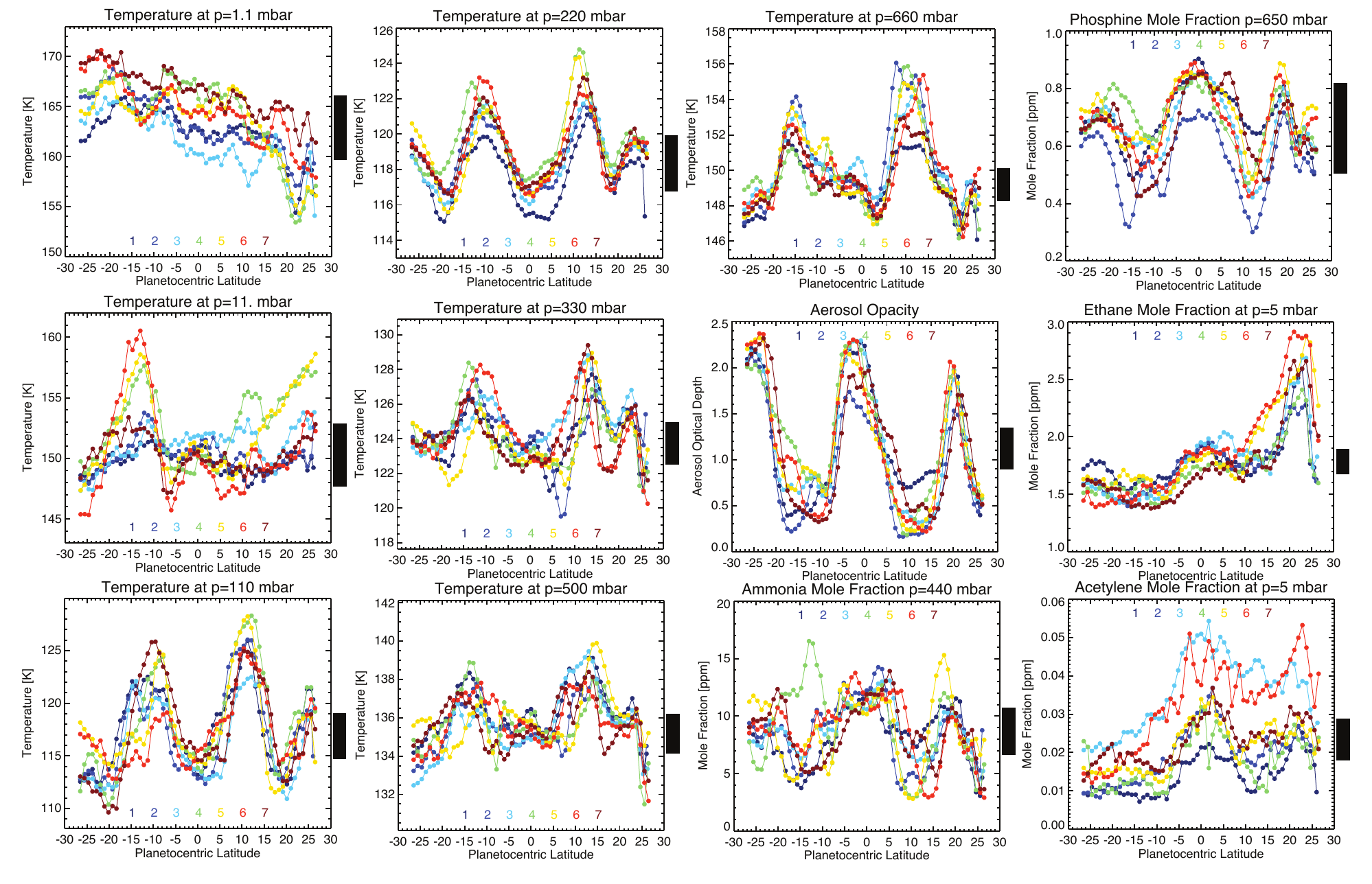}
\caption{Zonally-averaged Gemini/TEXES results for temperature, composition, and aerosol opacity. Each of the seven groups (March 12-14 2017) in Table \ref{tab:data} are included, as indicated by the key in each panel, each sampling a different central longitude.  The formal retrieval uncertainty is shown as the black bar to the right of each panel.  Latitudinal profiles at all pressure levels are available at \url{https://doi.org/10.5281/zenodo.3706796}. }
\label{zonal_texes}
\end{center}
\end{figure*}

With each TEXES group radiometrically scaled, we then proceed with a zonal-mean inversion of each group individually, deriving seven different latitudinal profiles for each parameter, in Fig. \ref{zonal_texes}.  Spectra for each group are binned on a $1^\circ$ latitude grid, retaining only those spectra within $10^\circ$ of the minimum emission angle for each latitude\footnote{Note that we omitted the 587-cm$^{-1}$ setting from the fitting due to excessive telluric water contamination.}.  A number of tests were performed to decide how to handle the prior imposed by the MWR profiles of deep NH$_3$ from \citeA{17li}: (i) fixing the deep NH$_3$ to a latitudinally-uniform mean of the MWR profiles and allowing the abundance to vary for $p<800$ mbar; (ii) fixing the deep NH$_3$ to the latitudinally-varying MWR profiles and allowing the abundance to vary for $p<800$ mbar; (iii) simply using the mean MWR profile as a prior and allowing the abundance to vary both above and below $p=800$ mbar.  In the first and second case, we found that the retrieved upper-tropospheric temperatures became extremely cold, as this was required to reproduce the deep absorption features observed in the TEXES spectra with the low MWR-derived ammonia abundances in the upper troposphere.  Following similar experiments with Cassini/CIRS fitting (detailed in \ref{ref_model}), we elected to go with the third approach, assuming that NH$_3$ is well-mixed for $p>800$ mbar, and that PH$_3$ is well-mixed for $p>1$ bar \cite{09fletcher_ph3}, and allowing both the deep abundances and the fractional scale heights of each gas to vary. Although the resulting NH$_3$ distribution differs from that inferred from the MWR observations at $p>800$ mbar by \citeA{17li}, we find that this improved the quality of the TEXES fits, and makes little difference to the derived NH$_3$ abundance at the peak of the TEXES contribution functions. 

Scale factors were retrieved for the ethane, acetylene, and aerosol distributions, along with a full profile retrieval for $T(p)$.  Equilibrium para-H$_2$ is assumed everywhere, and we adopt a single compact cloud at $p=800$ mbar (assumed to comprise NH$_3$ ice crystals with a distribution of radii $r=10\pm5$ $\mu$m) to represent the cumulative aerosol opacity down to the 1-bar level. Despite the good vertical coverage of the TEXES contribution functions \cite<Fig. 7 of>{16fletcher_texes}, there exists a degeneracy between aerosols, temperatures, and ammonia in the 500-1000 mbar range, as deep as the NH$_3$ ice cloud.  This may be partially responsible for the anti-correlation of temperature and aerosol opacity in Fig. \ref{zonal_texes}, although we note that we cannot achieve an adequate fit to the data by assuming spatial uniformity in aerosols and varying temperatures alone.  Furthermore, the spatially-resolved retrievals in Section \ref{maps} show that our aerosol maps resemble visible-light images, and differ substantially from the temperature/ammonia maps.  This indicates that temperature, aerosol, and ammonia are indeed separable in TEXES inversions. 

\begin{figure*}
\includegraphics[angle=0,width=1.2\textwidth]{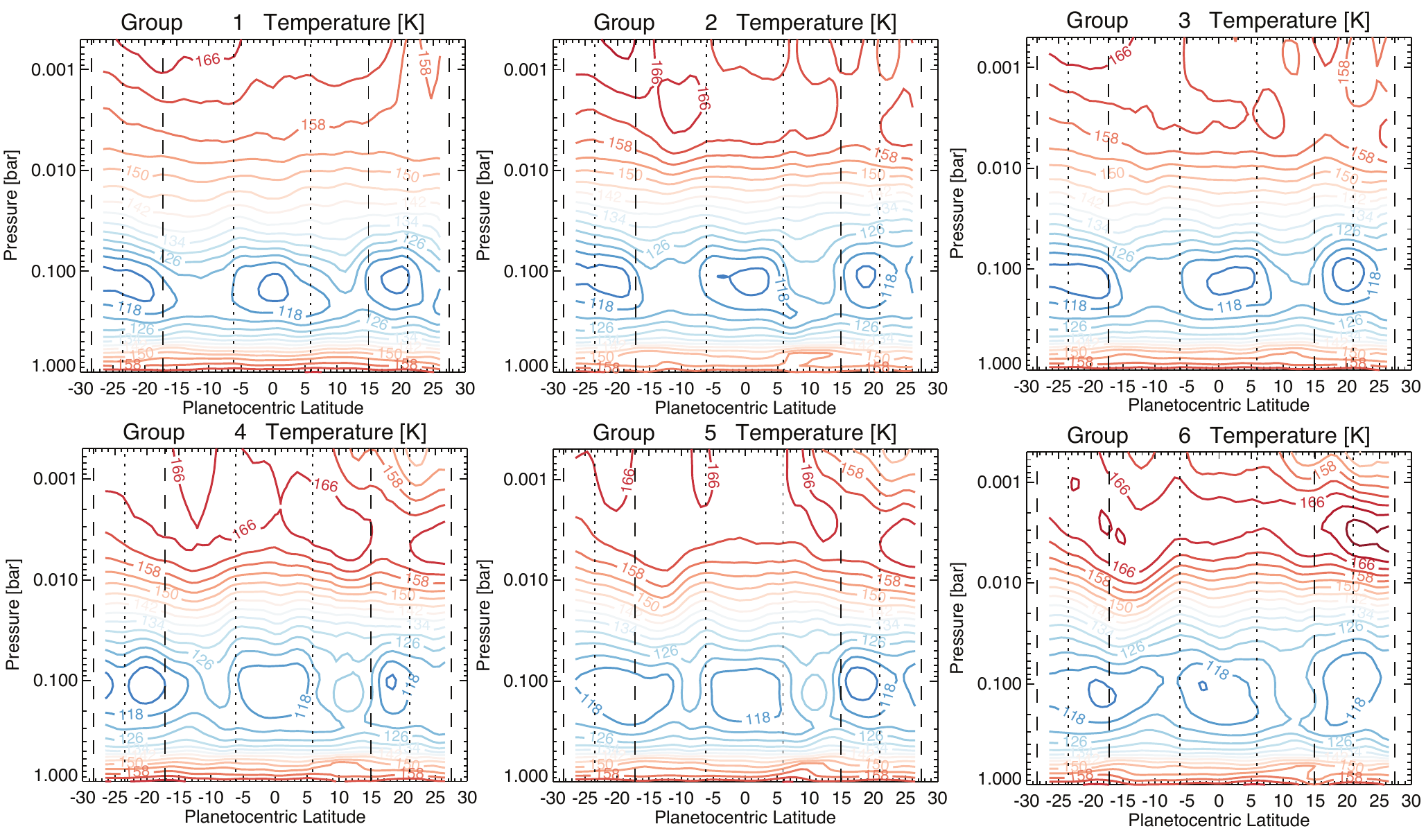}
\caption{Zonally-averaged Gemini/TEXES temperature contours, as a function of latitude and pressure. We include six of the seven groups in Table \ref{tab:data} (omitting group 7, as this is similar to all others), and the locations of Jupiter's prograde (dotted lines) and retrograde (dashed lines) jets are indicated, to show how they line up with the strongest $dT/dy$ gradients, and hence experience the strongest windshear with altitude.}
\label{zonal_temp}
\end{figure*}

\begin{figure*}
\includegraphics[angle=0,width=1.1\textwidth]{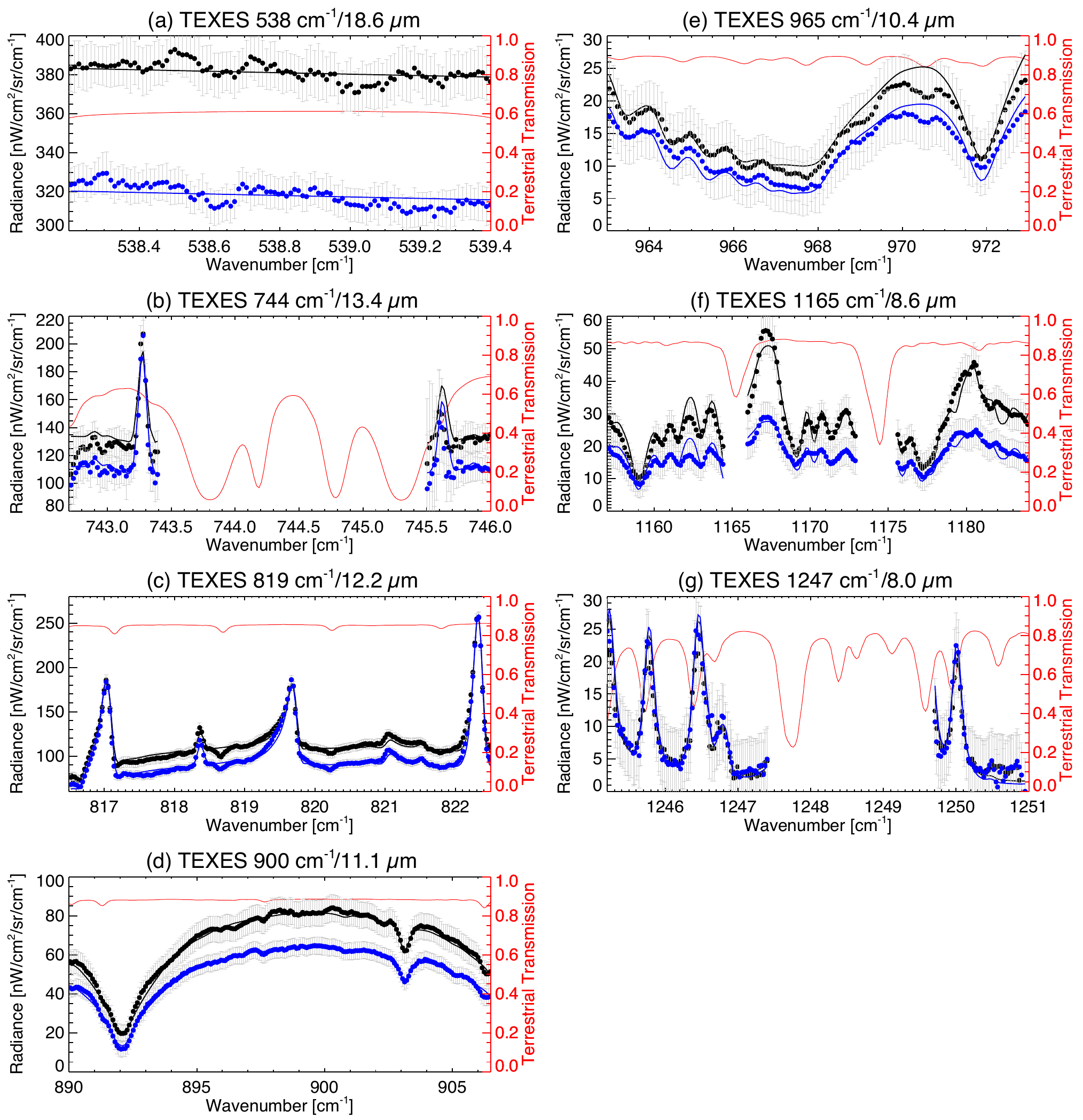}
\caption{The quality of the TEXES fits for group 1 in seven of the nine spectral channels (587 and 2145 cm$^{-1}$ were not used in the retrievals).  Data (points) and model fits (solid lines) are shown for the equator (blue) and for $10^\circ$N (black).  The red lines give the telluric transmission in each setting (registered to the right-hand axis), and serve as a guide to regions of the spectrum omitted from the fit.  Uncertainties (grey bars) and fitting procedures are described in \citeA{16fletcher_texes}.  }
\label{spectralfit}
\end{figure*}

Fig. \ref{zonal_texes} displays the latitudinal distributions of temperature, composition, and aerosols for each of the seven groups, and is supplemented by the temperature contours in Fig. \ref{zonal_temp}.  The key conclusion from this figure is that Jupiter's meridional profiles are highly variable as a function of longitude, such that if we only had a single group we would be misrepresenting Jupiter's true zonal average. The quality of the spectral fit is shown in Fig. \ref{spectralfit} for both the EZ and the NEB, where some regions of the spectra were omitted due to telluric contamination.  The retrieved parameters share much in common with those of \citeA{16fletcher_texes} and are briefly discussed here (stratospheric temperatures and hydrocarbon results are mentioned briefly in \ref{app_stratos}).  Tropospheric temperatures display a cool equator, warm NEB and SEB, adjacent cool zones at the NTrZ (North Tropical Zone) and STrZ (South Tropical Zone), and the warm belt of the NTB (North Temperate Belt).  These $dT/dy$ gradients, where $y$ is the north-south distance, are co-located with the peaks of the prograde and retrograde winds in Fig. \ref{zonal_temp}, implying a decay in the speed of the tropospheric jets into the upper troposphere and lower stratosphere.  The temperatures of the NEB and SEB are not symmetric, and vary strongly with longitude, implying that thermal windshears (and thus the speed of the winds at different altitudes) will also be longitudinally variable.  Ammonia, phosphine, and aerosol opacity are all elevated over the zones (EZ, STrZ and NTrZ) and depleted over the belts (NEB/SEB, and the NTB), supporting the canonical view of upwelling (and adiabatic cooling) in zones, and subsidence (and adiabatic warming) in belts.  The magnitude of the contrast depends strongly on the longitude, as expected from the influence of discrete features (e.g., plumes and hot spots) on the distribution of these species, necessitating the longitudinally-resolved retrievals in the following sections.

Figs. \ref{zonal_texes} and \ref{zonal_temp} clearly demonstrate the strong dependence of the retrieved temperatures, gases, and aerosols on the longitude.  This has significant consequences for estimates of the static stability, thermal winds, and the meridional gradients of potential vorticity, which are described in \ref{windshear}.  These products are derived from our retrieved temperatures for each TEXES group, and are made available in our Supplementary Material (\url{https://doi.org/10.5281/zenodo.3706796}) as a constraint on future modelling of Jupiter's tropics.  

\section{TEXES Tropospheric Maps}
\label{maps}

In this section we apply the spectral inversion technique of Section \ref{texes} to each of the seven groups to map NEBs features between $4^\circ$S and $17^\circ$N.  Each figure in the following sections spans $50^\circ$ of longitude, sufficient to capture 2-3 dark formations (DFs) or plume (PL) features on the NEBs jet at $6.0^\circ$N.  Using the tracking in System I longitude in Fig. \ref{jupos}, we match each group to a corresponding Juno perijove pass.  Despite the significant time that may have elapsed between a specific perijove and the March-2017 TEXES observations, almost all the DFs lasted throughout the period covered here (December 2016 to September 2017, PJ3 to PJ8), and were tracked showing slow westward drifts in System I (Figs. \ref{jupos}-\ref{irtf}).  The longevity of the DFs and plumes allows us to compare the TEXES maps with Juno maps at the time of the perijove, but we caution that the DFs could have evolved and changed during the interval.  We depict temperatures, phosphine, and ammonia at discrete levels - the $T(p)$ from a retrieval at all levels in our model, the gases from parameterised vertical profiles (a well-mixed abundance at high pressures, and a fractional scale height representing the decrease with altitude at lower pressures).  Full vertical profiles, plus stratospheric temperatures, ethane, and acetylene, are available in our Supplementary Material (\url{https://doi.org/10.5281/zenodo.3706796}).

\subsection{Perijove 4 Region - Group 4}

\begin{figure*}
\includegraphics[angle=0,width=1.0\textwidth]{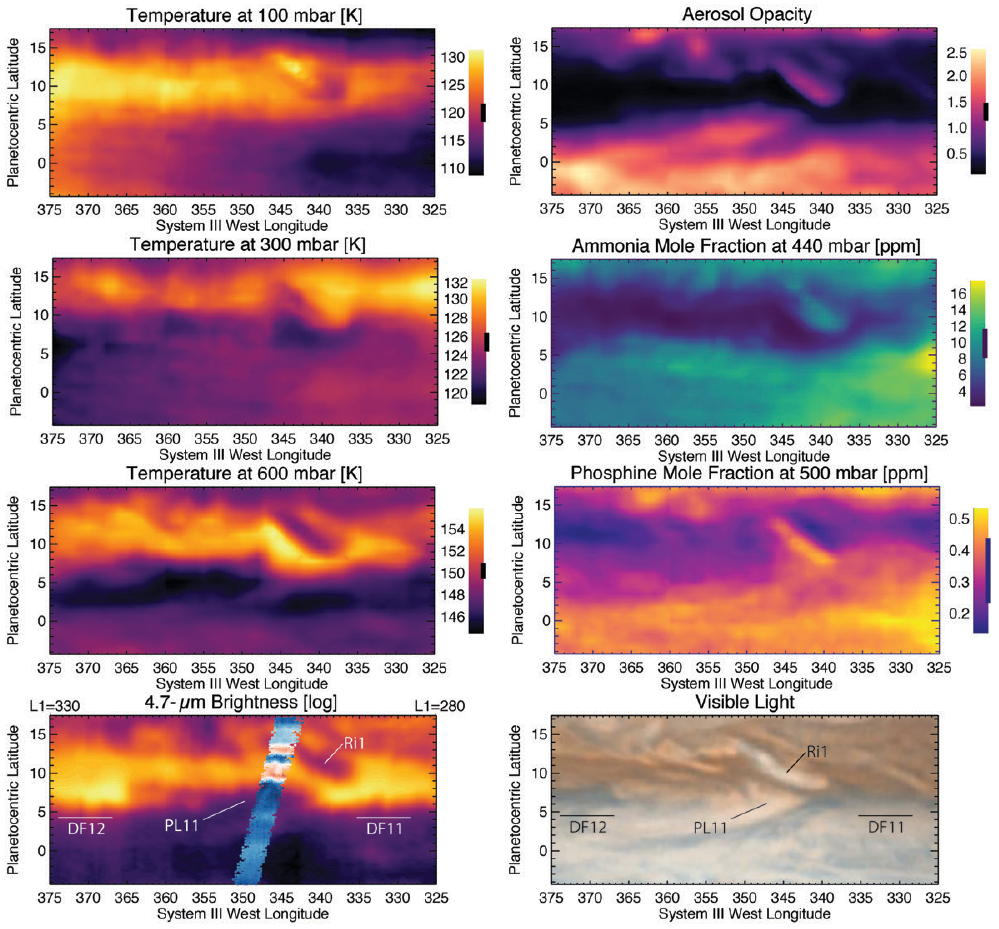}
\caption{Retrieved upper-tropospheric properties from TEXES group 4, 13 March 2017.  Temperatures at 100, 300, and 600 mbar are shown on the left, compared to the brightness at 4.7 $\mu$m (bottom left), which has been stretched logarithmically to reveal fainter features.  Retrieved aerosol opacity (cumulative optical depth to the 1-bar level) and ammonia and phosphine mole fractions are shown on the right, with a visible-light image from 20 hours later (D. Peach, 14 March 2017 at 06:24UT, adjusted in longitude to co-align the NEBs features).  We label the two dark formations (DF11/12), plume (PL11), and prominent NEB rift (Ri1).  Black vertical lines on the legend for each figure show the formal retrieval uncertainty.  The MWR PJ4 channel-6 (1.37 cm) brightness temperature from 02 February 2017 has been superimposed onto the 4.7-$\mu$m map to provide a qualitative comparison, co-aligned via conversion of the System-I longitudes from Fig. \ref{jupos}.  The System-I longitude range (L1) is added to the 4.7-$\mu$m map as a guide.}
\label{nebg4}
\end{figure*}

The potential plume observed by MWR (Fig. \ref{mwr_img}) and JIRAM (Fig. \ref{mwr_jiram}) on 02 February 2017 was observed in TEXES group 4 on 13 March 2017, 39 days later, near $345-355^\circ$W. In Fig. \ref{nebg4} we compare this to a visible-light image from Damian Peach taken $\sim20$ hours later, where the plume is labelled PL11, and is in between two dark formations, DF11 and DF12.  A prominent dark festoon extends southwest out of DF11 towards the equator, and borders the southeastern edge of PL11.  The plume is dark at 4.7 $\mu$m, partially due to the increased 800-mbar aerosol opacity, but potentially also due to the cooler physical temperatures at 600 mbar as shown Fig. \ref{nebg4}.  The festoon is moderately brighter at 4.7 $\mu$m due to aerosol depletion, consistent with the lower reflectivity of this feature in the visible-light map.  The dark formations are bright at 4.7 $\mu$m (consistent with their definitions as hot spots), but the physical temperatures of DF11 and DF12 at 600 mbar are not notably different from their surroundings, consistent with studies of the Galileo probe entry location \cite{98orton}.  Indeed, the warmest physical temperatures appear southwest of the prominent bright rift (Ri1 in Fig. \ref{nebg4}) in the NEB.  The rift itself is cold, reflective (white), and enhanced in ammonia, phosphine, and aerosols compared to the surrounding NEB, consistent with a vigorous upwelling from within the dry and depleted belt.  

The distributions of phosphine and ammonia are both spatially variable, but show neither strong depletion in DF11 and DF12, nor strong enrichment in PL11.  These features only appear readily in the physical temperature maps (and vary significantly with altitude) and in the aerosol map, suggesting that temperature and aerosol contrasts dominate the appearances of the plumes and dark formations, rather than the distributions of tropospheric gases.  Indeed, the most depleted NH$_3$ is located immediately south of the bright rift, which was not present during Juno's PJ4 (see Fig. \ref{mwr_jiram}).  The MWR Channel-6 brightness scan from 39 days earlier is approximately placed over PL11 in Fig. \ref{nebg4} to show that the coldest emission does co-align with the cold and aerosol-enriched plume, but shows less correspondence with the ammonia distribution near 440 mbar.  Thus far, we might conclude that ammonia gas is not playing a major role in the contrasts observed in MWR channel 6, which is more strongly affected by physical temperature.

\subsection{Perijove 5 and 6 Regions - Group 3}

\begin{figure*}
\includegraphics[angle=0,width=1.0\textwidth]{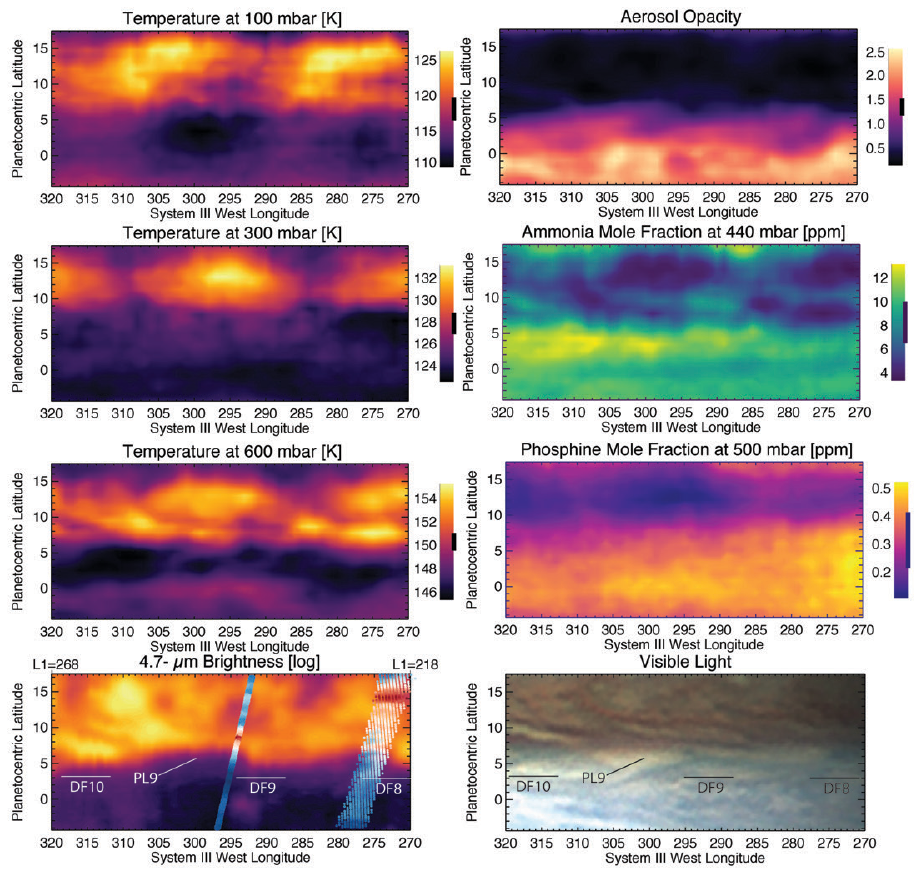}
\caption{Retrieved upper-tropospheric properties from TEXES group 3, 12 March 2017.  Temperatures at 100, 300, and 600 mbar are shown on the left, compared to the brightness at 4.7 $\mu$m (bottom left), which has been stretched logarithmically to reveal fainter features.  Retrieved aerosol opacity (cumulative optical depth to the 1-bar level) and ammonia and phosphine mole fractions are shown on the right, with a visible-light image from 10 hours earlier (A. Garbelini Jr., 12 March 2017 at 04:03UT, adjusted in longitude to co-align the NEBs features).  We label the three dark formations (DF8, 9 and 10), and a plume (PL9).  Black vertical lines on the legend for each figure show the formal retrieval uncertainty.  The MWR PJ5 (narrow) and PJ6 (broad) channel-6 (1.37 cm) brightness temperature have been superimposed onto the 4.7-$\mu$m map to provide a qualitative comparison (using the System-I longitudes from Fig. \ref{jupos}), but were taken 15-68 days after the TEXES observations.  The System-I longitude range (L1) is added to the 4.7-$\mu$m map as a guide.}
\label{nebg3}
\end{figure*}

The PJ5 MWR footprint on 27 March 2017 was very narrow in longitude (Fig. \ref{mwr_img}), and unlikely to have encountered either a plume or hot spot.  Nevertheless, a warm region associated with cloud-free NEB streaks was encountered, and TEXES observed this region on 12 March 2017, 15 days earlier, in group 3.  The PJ6 footprint on 19 May 2017 occurred just to the west of a prominent hot spot (DF8), which was also covered in group 3 on 12 March 2017, 68 days before PJ6.  The visible-light image in Fig. \ref{nebg3} suggests that the dark formations are not particularly prominent in this longitude range - once again, DF8-10 appear bright at 4.7 $\mu$m, with another plume (PL9) appearing dark due to excess aerosol opacity (top right of Fig. \ref{nebg3}).  And once again the physical temperature contrasts associated with the plumes and hot spots are extremely subtle, with the coldest 600-mbar equatorial features not necessarily co-located with structures at 4.7 $\mu$m nor visible wavelengths.  At lower pressures (100 and 300 mbar) we see evidence for a mid-NEB thermal wave in the temperature field \cite<as detailed in>{17fletcher_neb} that was not evident in Fig. \ref{nebg4}.   

As in the previous example, the hot spots do not notably perturb the ammonia and phosphine distributions.  However, the distribution of ammonia is intriguing:  the highest abundances are present in the $3-5^\circ$N range and may be associated with the most reflective features in visible light (e.g., white clouds, particularly near $300-310^\circ$W).  Rifts of more reflective material in the NEB near $10-12^\circ$N exhibit elevated NH$_3$ compared to the surroundings.  Superimposing the PJ5 and PJ6 MWR channel-6 brightnesses onto the retrieved maps suggest that both physical temperature contrasts and ammonia gas contrasts could be modulating the MWR brightness - for example the brightest 1.37-emission at $9^\circ$N (PJ5) and $14^\circ$N (PJ6) coincide with both warm temperatures at 600 mbar and locations of NH$_3$ depletion within the NEB.  In summary, the dark formations in this longitude domain are aerosol-depleted, but have limited signatures in the temperature and gaseous distributions.  Conversely, the brightest clouds within the plumes and rifts within the NEB show both temperature and ammonia contrasts that could be modulating the MWR brightness scans.

\subsection{Perijove 7 Region - Group 7}

\begin{figure*}
\includegraphics[angle=0,width=1.0\textwidth]{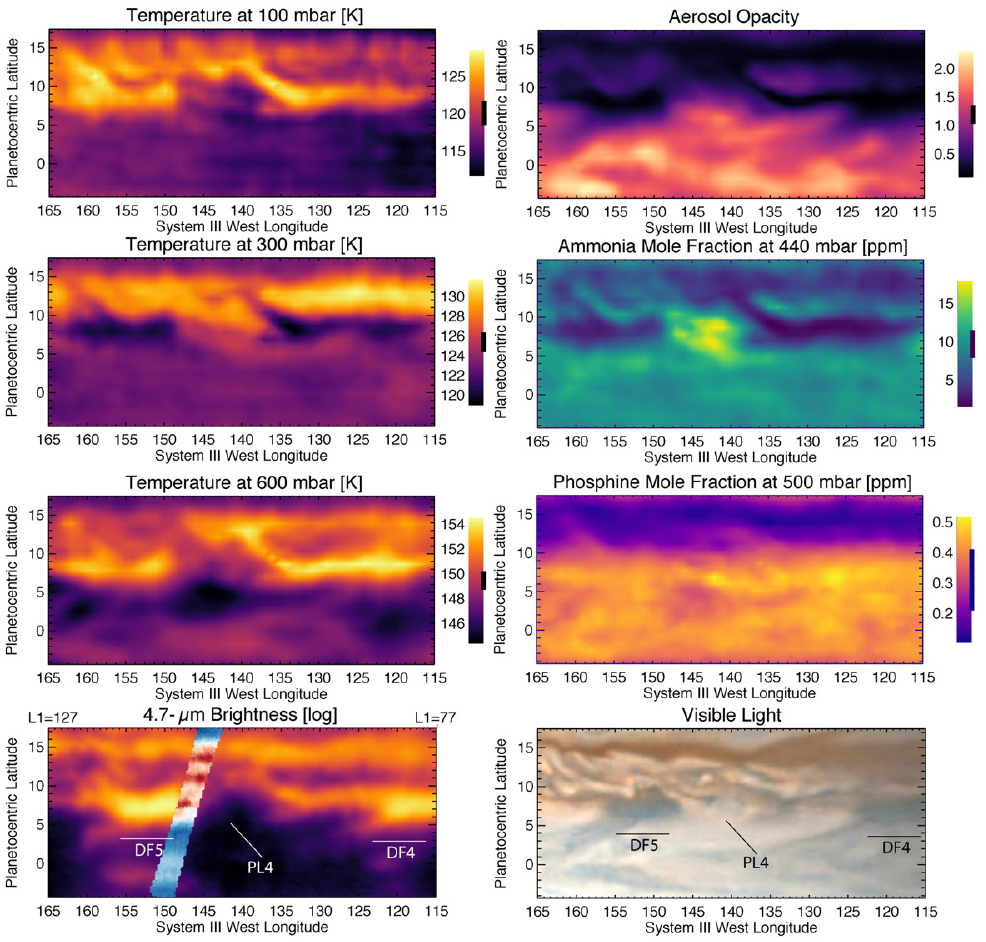}
\caption{Retrieved upper-tropospheric properties from TEXES group 7, 14 March 2017.  Temperatures at 100, 300, and 600 mbar are shown on the left, compared to the brightness at 4.7 $\mu$m (bottom left), which has been stretched logarithmically to reveal fainter features.  Retrieved aerosol opacity (cumulative optical depth to the 1-bar level) and ammonia and phosphine mole fractions are shown on the right, with a visible-light image from 20 hours later (D. Peach on 15 March 2017, 07:13UT, adjusted in longitude to co-align the NEBs features).  We label the two dark formations (DF4 and 5) and a plume (PL4).  Black vertical lines on the legend for each figure show the formal retrieval uncertainty.  The MWR PJ7 channel-6 (1.37 cm) brightness temperature has been superimposed onto the 4.7-$\mu$m map to provide a qualitative comparison (using the System-I longitudes from Fig. \ref{jupos}), but this was taken 119 days after the TEXES observations.  The System-I longitude range (L1) is added to the 4.7-$\mu$m map as a guide.}
\label{nebg7}
\end{figure*}

The PJ7 pass on 11 July 2017 provided MWR's closest encounter with an NEBs hot spot in this study (Fig. \ref{mwr_jiram}), although the MWR brightness was similar to the rest of the NEB in Channel 6 (sensing 700 mbar), but displayed a higher contrast in Channel 5 (sensing 1.5 bar).  Given that 119 days had elapsed, the PJ7 region ($L1=167.4^\circ$) would have been located near $L1=110-115^\circ$W in March 2017 (the hot spots drift westward in System-I longitude, Fig. \ref{jupos}), which was covered by TEXES group 7 in the $L3=147-153^\circ$W region, labelled as DF5 in Fig. \ref{nebg7}.  We see the familiar pattern - DF4 and DF5 are bright at 5 $\mu$m and depleted in aerosols at $\sim800$ mbar, physical temperatures at 600 mbar are elevated, NH$_3$ is generally depleted within the hot spot, although no strong signatures are observed in the distribution of phosphine.  In this instance, the temperature and ammonia contrasts are more closely correlated to the aerosol distributions, indicating that each of the hot spots is different.  Indeed, the hot spots appear to have extensions that penetrate further north into the NEB, mixing with the rifting regions.

Fig. \ref{nebg7} is most remarkable for the plume of enhanced NH$_3$ between $140-145^\circ$W and $6-10^\circ$N, co-located with enhanced aerosol opacity and low 5-$\mu$m brightness.  This plume (PL4) appears as a region of bright, reflective clouds in the visible-light image acquired in March 2017.  The ammonia enrichment is higher here than over the rest of the EZ, and is most similar to the MWR scan during PJ4, where low Channel-6 brightness temperatures were observed up to $8^\circ$N.  Compared with the observations throughout this section, it seems that plumes are sometimes (but not always) elevated in ammonia gas, and that each plume is different, just like the hot spots.

\subsection{Perijove 3 and 8 Regions - Group 6}



\begin{figure*}
\includegraphics[angle=0,width=1.0\textwidth]{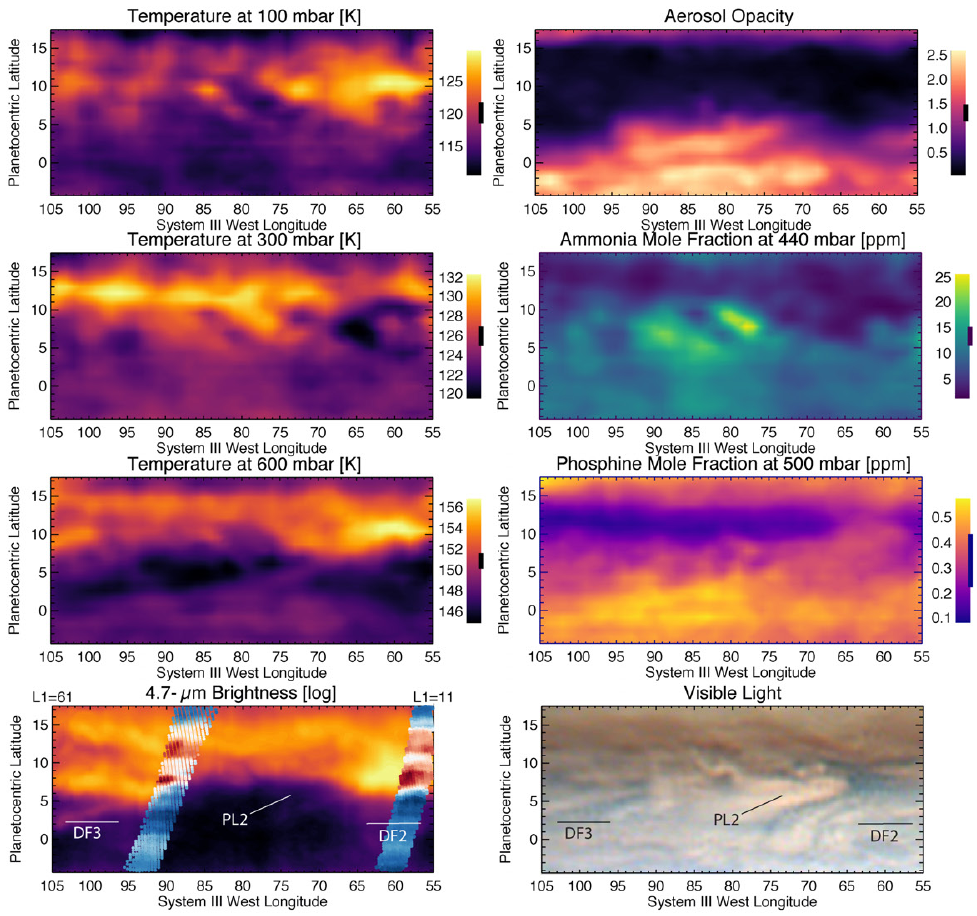}
\caption{Retrieved upper-tropospheric properties from TEXES group 6, 13 March 2017.  Temperatures at 100, 300, and 600 mbar are shown on the left, compared to the brightness at 4.7 $\mu$m (bottom left), which has been stretched logarithmically to reveal fainter features.  Retrieved aerosol opacity (cumulative optical depth to the 1-bar level) and ammonia and phosphine mole fractions are shown on the right, with a visible-light image from 2.5 days later (S. Kidd on 15 March 2017, 23:43UT, adjusted in longitude to co-align the NEBs features).  We label the TWO dark formations (DF2 and 3) and a plume (PL2).  Black vertical lines on the legend for each figure show the formal retrieval uncertainty.  The MWR PJ3/8 channel-6 (1.37 cm) brightness temperature has been superimposed onto the 4.7-$\mu$m map to provide a guide for DF2/3 (using the System-I longitudes from Fig. \ref{jupos}), respectively, but note the large time separation between the MWR and TEXES observations.  The System-I longitude range (L1) is added to the 4.7-$\mu$m map as a guide.}
\label{nebg6}
\end{figure*}

TEXES Group 6 (13 March 2017) covered DFs and plumes in the regions (in System I longitudes) observed during PJ3 (11 December 2016, 92 days before the TEXES maps) and PJ8 (01 September 2017, 172 days after the TEXES maps).  Table \ref{tab:pjs} gives the $L1$ longitude of each perijove, which we adjust for the drift of NEBs features in Fig. \ref{jupos}, and find that the putative PJ3 hot spot (DF2) should be near System-III longitudes of $60^\circ$W, and the PJ8 hot spot should be near $95^\circ$W (DF3). This is confirmed by the 5-$\mu$m brightness in Fig. \ref{nebg6}, although the PJ8 hot spot (DF3) is more elongated and dimmer than the more compact PJ3 hot spot (DF2).

Group 6 confirms the trends highlighted in the previous sections.  DFs are warmer and aerosol-depleted compared to the plumes, but whereas DF2 exhibits elevated 600-mbar temperatures and a spatially-complex NH$_3$ depletion, the temperatures and ammonia of DF3 were indistinguishable from their surroundings in March 2017, even though the DF was bright at 4.7 $\mu$m (e.g., Fig. \ref{irtf}).  The fact that PJ8 did observe enhanced 1.4-cm brightness (i.e., NH$_3$ depletion and/or an increased kinetic temperature) over the eastern edge of DF3 in September 2017 implies that the DF evolved in the intervening 6 months (i.e., the maturity of the DF influences the ammonia/temperature distributions).  The plume PL2 stands out in the NH$_3$, temperatures, and aerosol maps, with some regions of enhanced NH$_3$ and cold temperatures in the $5-10^\circ$N region between the two DFs.  Once again, the retrieved aerosol opacity most closely resembles the visible albedo and 5-$\mu$m brightness, but not the temperatures and NH$_3$ that are more important for the MWR observations.  Thus we should not expect microwave maps to always resemble observations that primarily sense aerosols (Fig. \ref{mwr_jiram} and \ref{junocam}).  Finally, the PH$_3$ is not perturbed by the hot spots and plumes, instead showing the regular enrichment over the EZ (and NTrZ at the top of the map) and depletion over the NEB.      

TEXES groups 1, 2 and 5 provided access to some additional prominent hot spots and plumes in March 2017, and the retrieved maps are available in the Supporting Information.  These confirm the DFs as complex objects, spatially inhomogeneous in temperature and ammonia from east to west, rather than being uniformly depleted in ammonia (even though aerosol depletion/enrichment seems to span the whole hot spot/plume feature, respectively).  The Juno observations and Galileo probe measurements therefore depend on \textit{where} in the DF/plume the observations took place, which points to the need for MWR scans with broader longitudinal coverage to capture the full extent of a dark formation.


\section{Forward Modelling MWR Contrasts}
\label{mwrmodel}

\begin{figure*}
\includegraphics[angle=0,width=1.2\textwidth]{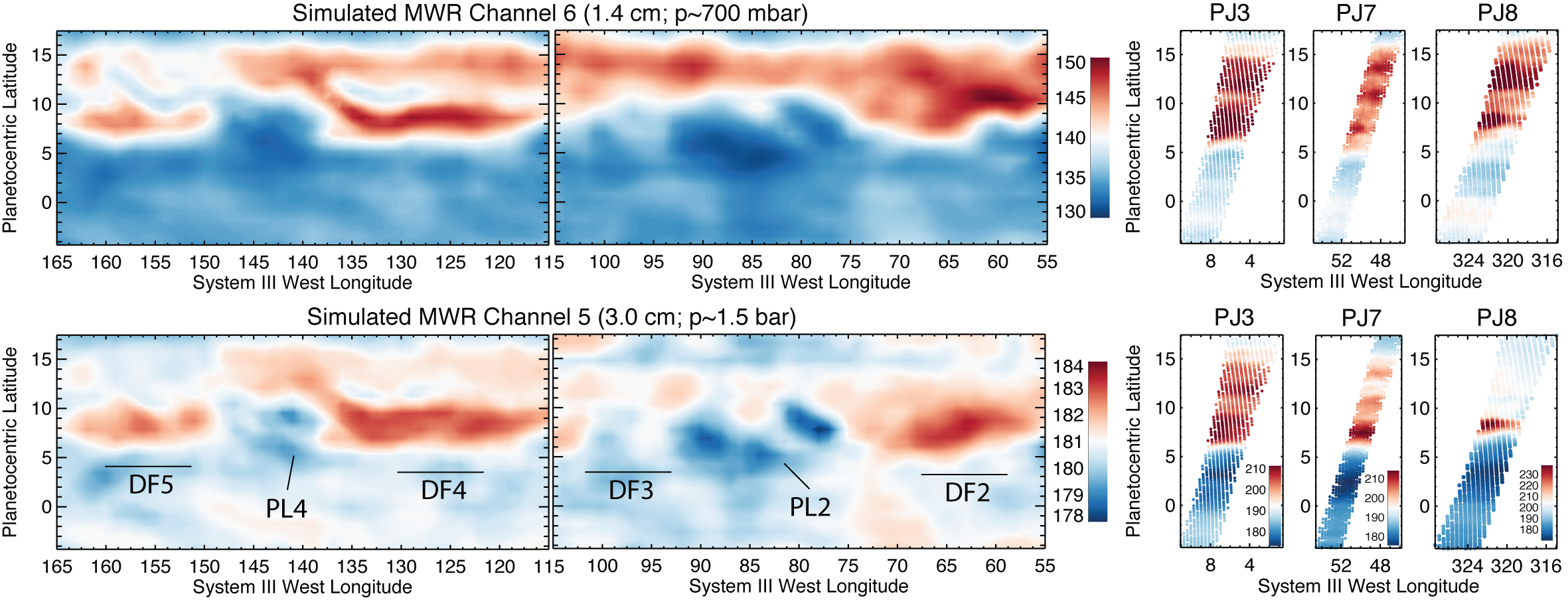}
\caption{Forward-modelled nadir MWR brightness in channel 5 (3.0 cm, bottom row) and 6 (1.4 cm, top row) based on the TEXES-derived temperatures and ammonia distributions from group 6 (right, Fig. \ref{nebg6}) and 7 (left, Fig. \ref{nebg7}).  The expected locations of four of the dark formations (DFs) are labelled.  These maps show what MWR might see if it could span the entire longitude domain.  Nadir-equivalent brightness scans during PJ3, 7 and 8 are shown on the right as examples of the real data, though note that these are separated from the TEXES maps by several months.  They are plotted on the same colour scale as the models for Channel 6, but different colour scales for Channel 5 where the model is a poor reproduction of the data.  }
\label{fmodel}
\end{figure*}

We can use the TEXES-derived ammonia and temperature distributions (sensing $p<800$ mbar) to understand whether the ground-based observations can successfully predict microwave brightness variations observed by MWR in channels 5 ($p\sim1.5$ bar) and 6 ($p\sim700$ mbar).  The forward model of \citeA{17li} was used to simulate the expected MWR nadir brightness temperatures based on the TEXES results for groups 6 and 7, spanning just over $100^\circ$ of longitude and covering at least four DFs (hot spots) and two plumes (PL2 and PL4) in Fig. \ref{fmodel}.  Despite the separation in time between the March 2017 TEXES observations and the Juno MWR maps, we also show the channels 5 and 6 nadir-equivalent brightness temperatures for PJ3 (December 2016), 7 (July 2017) and 8 (September 2017) for comparison.  Channel 6 (1.4 cm), which sounds pressures near the 700-mbar level, is reasonably well-reproduced by the TEXES predictions, allowing for some systematic offsets (the model is too cool at the equator and at the brightest points in the NEB).   Not only does the model reproduce the contrasts associated with the DFs, but shows that we should expect low-amplitude brightness variability throughout the equatorial zone, which is consistent with the MWR observations.  

However, the bright features in channel 5 (3.0 cm) are poorly reproduced.  This is expected, as the information content for the TEXES spectra drops severely for pressures exceeding 800 mbar, meaning that our retrieved profiles relax to the assumed priors.  The model matches the coolest temperatures ($\sim180$ K) but fails to reach the 210-230 K brightness of the hot spots.  The DFs are either physically warmer or much more NH$_3$-depleted at 1.5 bar (or some combination of the two) than the TEXES inversions can account for.  The only thing that we do reproduce in Channel 5 is the fact that $6-9^\circ$N is the brightest region in this domain, whereas in Channel 6 the hot spots are similar to other bright features within the NEB.  We also note that one of the hot spots, DF3 (which was encountered during PJ8 in September 2017), does not show up in our predicted MWR maps for March 2017.  We know that the DF was present throughout the 2016-17 apparition (Fig. \ref{jupos}), so it cannot be regarded as particularly `new' in March 2017.  The NH$_3$ and temperature maps in Fig. \ref{nebg6} also failed to reveal any prominent contrasts there, which confirms two things: (i) the hot spot must have evolved significantly between March and September to create the bright emission observed by MWR; and (ii) each DF can show remarkably different gaseous and thermal contrasts, highlighting the need to observe them simultaneously from different facilities.  This also confirms that the 5-$\mu$m observations and visible reflectivity (most sensitive to aerosols) need not necessarily correlate spatially with the temperatures and gases sensed by TEXES and MWR.  Thus comparisons to aerosol maps (e.g., Fig. \ref{mwr_jiram} and Fig. \ref{junocam}) are not necessarily a good proxy for the expected microwave brightness, although they can certainly provide a qualitative guide.  

Ammonia-rich equatorial plumes are observable as the coldest features near $5-10^\circ$N, and are prominent in both channel 5 and 6.  Indeed, the low MWR brightness extends well into the NEB, as was seen during PJ4, which flew over part of one of these plumes.  The 20-K contrast between the plumes and the DFs is clear in Fig. \ref{fmodel}, and is responsible for the large variability from PJ to PJ that was observed in the zonally-averaged brightnesses in Fig. \ref{mwr_zonal}.  Finally, although we find broad consistency between TEXES and MWR in the upper troposphere (700 mbar), the next step in this research will be a joint inversion of mid-infrared and microwave observations to find the atmospheric structure consistent with the deeper-sounding MWR observations too.

\section{Discussion and Conclusions}
\label{discuss}

Gemini TEXES observations, combined with amateur tracking of the locations of NEBs plumes and hot spots, provide contextual maps of temperature, aerosols, ammonia and phosphine surrounding features observed by Juno's MWR, JIRAM, and JunoCam instruments in Jupiter's tropics.  To our knowledge, these are the highest-resolution thermal-infrared spectral maps of Jupiter's equatorial zone and belts, and the first to reveal inhomogeneities \textbf{within} the hot spots themselves.  We summarise the results section as follows:

\begin{enumerate}
    \item \textbf{Aerosols:} The defining characteristic of the NEBs features is their aerosol content:  bright/dark regions at 5 $\mu$m correlate extremely well with the absence/presence of aerosol opacity inferred from spectra near 8.6 $\mu$m (1165 cm$^{-1}$), and with the visible reflectivity (e.g., see in particular Figs. \ref{nebg7} and \ref{nebg6}).  The trapped equatorial Rossby wave appears to primarily modulate the aerosols, which explains why JIRAM 5-$\mu$m maps and JunoCam visible-light maps match the derived aerosols fields so well.  However, aerosol maps are not a good proxy for microwave maps \cite<as previously suggested by>{17orton}, which are primarily sensitive to temperature and ammonia.
    \item \textbf{Temperature and Ammonia: } The dark formations are not uniformly warmer than their surroundings, nor are they uniformly depleted in NH$_3$ gas in the 500-700 mbar range.  Some of the DFs appear to have longitudinal inhomogeneities in temperature and ammonia content, implying that the microwave brightness sensed by MWR is extremely sensitive to regional variations within the hot spots.  The same is true of the plumes - they are not uniformly colder than their surroundings, and they are not uniformly enriched in NH$_3$.  In one extreme case (DF6), the whole eastern edge of a DF was hidden by a region of enriched NH$_3$ gas, stretching from the northern EZ into the NEB.  This shows the need for broader longitudinal coverage in MWR scans to understand these features.
    \item \textbf{Low Microwave Contrasts at 700 mbar:} The MWR-derived brightness of dark formations at 1.37 cm (Channel 6, sensing contrasts in both ammonia and temperature) is not noticeably different from other bright features within the NEB.  This is supported by the TEXES-derived ammonia maps, which often show only subtle contrasts between the NEB and the DFs/plumes.  This lack of contrast implies that the \textit{in situ} ammonia and temperature measurements from the Galileo probe, at least at the $p<1$ bar levels within the hot spot, may have been more representative of the wider NEB than previously thought.
    \item \textbf{Depth of hot spots: } NEBs structures appeared to show the largest contrasts in MWR Channel 5 (3.0 cm, sounding 1.5 bar), where the hot spots DF3 and DF5 provided high microwave brightness.  Signatures could still be observed in Channel 4 (5.75 cm, sounding $\sim3.5$ bar), where brightness temperatures are comparable to those measured at 5 $\mu$m, but by Channel 3 (11.55 cm, sensing 10 bar) the hot spots were indistinguishable from their surroundings.  This implies that any ammonia (or water) depletions associated with the DFs are a feature of the atmosphere above the $p=10$-bar level, rather than extending into the deeper atmosphere.  This also implies that the volatile depletions observed by the Galileo-probe and Juno measurements for higher pressures are representative of the entire NEB, rather than being unique conditions due to DF meteorology.
    \item \textbf{Absence of PH$_3$ Contrast:  } The TEXES dataset does include some notable examples of elevated PH$_3$ associated with vigorous plumes in the NEB and SEB, suggesting that these rise from sufficient depths to bring PH$_3$ upwards.  However, the dark formations and equatorial plumes do not display contrasts in the PH$_3$ distribution.  This may be further evidence that the NEBs features are relatively `shallow' phenomena, having little impact on the distribution of PH$_3$.  
    \item \textbf{Differences between hot spots: } The TEXES maps examined the conditions within all 12 of the dark formations present in March 2017.  In some cases, warm temperatures and depleted NH$_3$ were coincident with the cloud-free conditions, but in other cases there were only subtle contrasts observed.  No two hot spots or plumes were equivalent (see, for example, the different morphologies of DFs at 4.7 and 8.6 $\mu$m in Fig. \ref{map1165}), perhaps reflecting evolution during their multi-month lifetimes (e.g., a sign of their `maturity'), or reflecting interactions of the NEBs features with the surrounding atmosphere.  This may point to the uniqueness of the Galileo-probe measurements (and some of the Juno MWR scans), at least for $p<10$ bars.  Extending this survey over the multi-year Juno mission would help to disentangle these effects.
\end{enumerate}

\subsection{Static Stability in Hot Spots}
The TEXES inversions showed one feature that we have not yet adequately explained - for the hot spots themselves, we see a `kink' in the temperature profile near 700-800 mbar, represented as a distinct change in the lapse rate $dT/dz$.  The typical retrieved lapse rate is $\sim1.6$ K/km near 600 mbar (i.e., about 0.5 K/km smaller than the dry adiabatic lapse rate of 2.1 K/km), decreasing to $\sim1.0$ K/km at 300 mbar.  However, the hot spots displayed 600-mbar lapse rates approaching $\sim2.5$ K/km (i.e., 0.4 K/km larger than the dry adiabatic lapse rate, and therefore unstable), which causes the temperatures to quickly go from warmer than the surroundings at 600 mbar to cooler than the surroundings near 300 mbar.  At higher pressures ($p>700$ mbar), $dT/dz$ decreased to 0.5-1.0 K/km (i.e., a quasi-isothermal layer exists near 700-900 mbar, potentially as a result of volatile condensation releasing latent heat to increase the static stability).  This causes the 1-bar temperatures to be 2-3 K cooler than the Galileo-probe derived value of 165 K \cite{98seiff}, although we caution that this is at the limit of the retrieval accuracy.  Comparing the Brunt V\"{a}is\"{a}l\"{a} buoyancy frequency $N$ (a measure of the stratification, see \ref{windshear}) between hot spots and their surroundings, we find that unstable ($N^2<0$) regions exist in the 450-700 mbar region near the NEBs, but that this extends slightly deeper (to 800-900 mbar, near the limit of our vertical sensitivity) within the hot spots.  The hot spots were the only regions in the entire TEXES maps that displayed this unusual $dT/dz$ and $N$, suggesting that this is not a general artefact of the retrieval process.  

We explored numerous potential sources for this phenomenon in our retrievals - our assumptions for vertical parameterisations of PH$_3$, NH$_3$, and aerosols; potential temperature-composition degeneracies; the degree of vertical smoothing applied in the retrieval process; we even attempted to remove aerosol opacity from the hot spots entirely (this prevents us from achieving an adequate fit).  Unfortunately neither Voyager IRIS nor Cassini CIRS has the spatial resolution to resolve these features, so we were unable to validate our cross-calibration for the hot spots, although all other regions of Jupiter seemed to be fitted very well.  We conclude that these unstable $dT/dz$ profiles are a real, if unexplained, feature of the hot spots themselves.  

To examine the validity of this high lapse rate, we looked to Galileo Probe measurements.  The atmospheric structure instrument (ASI) featured different pressure sensors that operated in different regimes \cite{98seiff}:  the $p_3$ sensor (reliable for $p>4$ bar) recorded questionable spikes in lapse rate up to 3 K/km near 0.5 bar, but they were not seen by the $p_2$ sensor (reliable for $p<4$ bar).  Although the latter saw lapse rates approaching 2.5 K/km near 1.7 bar, the validity of these unstable layers was questioned by \citeA{98seiff}. Furthermore, using the ASI temperature sensors instead of the pressure sensors, \citeA{02magalhaes} suggested that the Galileo-probe profile was actually stable for $p<10$ bar, with small deviations from the adiabat of only 0.1-0.2 K/km in the 0.5-1.7 bar region, which they tentatively associated with heating due to particulates and aerosols observed by the nephelometer \cite{98ragent} and net-flux radiometer \cite{98sromovsky}.  Maybe the `kink' we see near 700-900 mbar is associated with the more stable regions of the profile found by \citeA{02magalhaes}.  However, for the Galileo hot spot in 1995, the existence of unstable layers seems doubtful.  Nevertheless, this does not rule out the possibility of unstable layers at $p<700$ mbar within the 2017 hot spots explored by TEXES.  The absence of aerosols in the column above the hot spot might serve to preferentially cool the atmosphere via radiative emission; or the visibly-dark formations might absorb more heat than the more reflective surroundings.  Both processes might contribute to destabilising the atmosphere within one of these features, but we might expect this to increase the vigour of mixing, which is not seen within these DFs even with JunoCam imaging (Fig. \ref{junocam}).  Further progress on precisely measuring the $dT/dz$ may be made via better-calibrated space-based mid-IR spectroscopy of hot spots and their surroundings from the James Webb Space Telescope in the coming decade.


\subsection{Depth of Hot Spots and Plumes}
As discussed in Section \ref{intro}, conditions within the equatorial Rossby wave can be represented via downward stretching of an atmospheric column within the hot spots \cite{00showman, 05friedson}.  This serves to distort material surfaces towards higher pressures, explaining why neither the Galileo-probe measurements of the condensable gases \cite{04wong} nor the cloud bases \cite{98sromovsky} were in the locations predicted by equilibrium condensation models.  Indeed, the Galileo probe measurements showed that ammonia began to reach a uniform abundance near 8-10 bars \cite{04wong_gal, 19depater_vla}, and H$_2$S near 16 bars \cite{98niemann, 98folkner}.  The implied downwelling also forces the clouds to sublimate, rendering the DFs bright at 5 $\mu$m.  Recently, \citeA{18li_adiabat} showed how a single stretch parameter approximately reproduced the probe measurements, moving the transition pressure (where the abundance becomes approximately well-mixed with altitude) for NH$_3$ from 1 bar down to 3-4 bars, and for H$_2$S to 10 bars.  For NH$_3$, this is consistent with a lack of microwave signatures of the hot spots for $p>10$ bar \cite<modelling work by>[places the ammonia transition near 8 bars]{19depater_vla}, as described above and shown in Fig. \ref{mwr_zonal}.  




A pattern is emerging that suggests that the equatorial Rossby wave may be confined to $p<5-10$ bars, and that it can be considered as a `weather-layer' phenomenon primarily affecting ammonia abundances (and possibly H$_2$S), cloud opacities, and local temperatures.   This is consistent with the absence of plume/hot spot contrasts in the microwave brightness at higher pressures (i.e., no NH$_3$ or H$_2$O contrasts in MWR channels 1-3), and the absence of strong contrasts in PH$_3$ derived in the upper troposphere.  Thus Galileo-probe measurements for higher pressures (where winds became stable, ammonia well mixed, but water remained depleted) could be representative of the entire NEB for $p>5-10$ bar, rather than a consequence of the unusual conditions within the hot spot.  However, one would still have to explain the apparent depletion of H$_2$O across the NEB, which is also the site of localised lightning events \cite{00gierasch, 00ingersoll}.  Resolving this conundrum requires numerical simulations with greater vertical resolution, combined with more comprehensive microwave measurements of mature hot spot features.  Thermal and compositional profiles produced by numerical simulations should be forward-modelled for direct comparison to mid-infrared and microwave observations.  Finally, we note that none of the features observed by MWR in 2017 reached the microwave brightness (and thus NH$_3$ depletion) that one might expect from the Galileo probe hot spot, and work is ongoing to correlate microwave brightness measured by Juno during the remaining years of its mission with dynamic phenomena observed in Jupiter's tropical domain.

\appendix
%
%

\section{TEXES image cube destriping}
\label{destripe}
The Gemini/TEXES slit is shorter (15") than is typically used on IRTF, meaning that the edge of the slit does not overlap with sky background to allow for removal of any temporal variations in the water column (and hence the sky background) between each step of the scan.  The resulting scans therefore exhibit striping in the direction parallel to Jupiter's rotation axis, with the settings with the worst telluric contamination (539, 587, 745 and 1248 cm$^{-1}$) exhibiting the strongest banding.  To correct this, the image at each wavelength was transformed into Fourier space, where the frequencies of the vertical striping could be identified and masked out before transforming back into the image domain.  The mask is applied as a thin horizontal bar near $y=0$, preserving all other spatial frequencies and making the assumption that only stripe artefacts would be correlated over all $y$-pixels for a particular location, $x$.  Note that this bar could not impinge too closely on the central burst (the lowest spatial frequencies), as this would start to remove the large-scale structure in the images.  For each setting, a balance had to be identified between removing the stripes and removing real jovian information.  The reconstructed images from the masked 2D Fourier spectra are prone to errors at the edges of the disc, so spectra at high emission angles are omitted from the this analysis.

\section{CIRS Radiometric Reference Model}
\label{ref_model}

In order to find a well-calibrated baseline for the TEXES spectra from the IRTF, \citeA{16fletcher_texes} retrieved temperatures, aerosols, phosphine, ammonia, acetylene and ethane from Cassini Composite Infrared Spectrometer (CIRS) observations of Jupiter in December 2000 - specifically the ATMOS02A 2.5-cm$^{-1}$ resolution map acquired on December 31st 2000 near closest approach to Jupiter.  The ammonia profile used as a prior for the CIRS inversions came from previous CIRS studies \cite{04fouchet, 06achterberg, 09fletcher_ph3}, which had, in turn, obtained priors from Voyager \cite{92griffith} and ISO \cite{00fouchet} measurements.  None of these were able to sample significantly below Jupiter's $\sim500-700$-mbar aerosols (expected to be NH$_3$ ice, contaminated by chemicals that serve as chromophores).  Juno arrived at Jupiter in July 2016 \cite{17bolton}, and the Microwave Radiometer \cite<MWR,>{17janssen} permitted zonally-averaged NH$_3$ measurements to much higher pressures \cite{17li}.  The microwave-derived NH$_3$ distributions from PJ1 (27 August 2016) \cite{17li} could therefore be used to provide updated priors for the CIRS retrieval analysis.

However, an immediate obstacle was identified, in that the retrievals of \citeA{17li} assumed an adiabatically cooled tropospheric temperature right to the top of their domain, rather than a radiatively controlled $T(p)$.  This meant that their assumed upper-tropospheric temperatures (100-800 mbar) were much cooler than reality.  When the CIRS spectral range (600-1400 cm$^{-1}$ for the purposes of this study) was modelled using the MWR-derived ammonia, it was clear that the CIRS data (sampling $p<800$ mbar) were not being adequately reproduced.  As a compromise, we adopt the deep NH$_3$ and $T(p)$ structure from \citeA{17li} in the $p>800$ mbar region, and fit a fractional scale height (i.e., the ratio of the ammonia scale height to the atmospheric scale height) to reproduce the CIRS spectra.  We found no significant differences if we used 700, 800, or 1000 mbar as the transition point. 

We conducted four specific experiments - (A) including the MWR NH$_3$ profiles for each latitude and not varying them during the fitting of temperature, aerosols, phosphine, ethane and acetylene; (B) fixing the deep abundance to a latitudinally-uniform 277 ppm everywhere for all $p>800$ mbar (an average of the MWR results at this altitude) and varying the NH$_3$ scale height (along with the other parameters listed above); (C) fixing the deep $p>800$ mbar abundance to the latitudinally-varying MWR NH$_3$ abundance at 800 mbar for each latitude, and again varying the NH$_3$ scale height; and (D) allowing both the deep NH$_3$ abundance, and the fractional scale height above the 800-mbar level, to vary during the fitting process.  For the latter three tests, the December-2000 CIRS spectra require approximately 10-12 ppm of NH$_3$ at 440 mbar at the equator, compared to 2-5 ppm in the NEB and SEB.  This contrast between the equator and neighbouring belts is measured irrespective of whether the deep NH$_3$ at $p>800$ mbar is variable, or uniform, or fixed to the MWR profiles \cite{06achterberg, 16fletcher_texes}.  Conversely, the August-2016 NH$_3$ profiles of \citeA{17li} have only $\sim3$ ppm at the equator at 440 mbar, and $\sim1.5-2.0$ ppm in the belts, meaning that they did not provide sufficient absorption to reproduce the CIRS spectra.  We noted a moderate improvement to the fitting quality in the NEB when we assumed a latitudinally-uniform $p>800$ mbar abundance (test B), as opposed to taking the exact values from \citeA{17li} (test C).  This is because the 800-mbar NH$_3$ abundances from MWR range from $\sim300$ ppm at the equator to $\sim30$ ppm in the NEB, a factor-of-ten depletion.  Although not strictly ruled out by the CIRS spectra, this low NEB abundance produced a poorer fit to the NEB than when we assumed 277 ppm at $p>800$ mbar.  We caution that we cannot rule out a real temporal change in the NEB NH$_3$ content between December 2000 and August 2016.

Test D then ignored the MWR abundances entirely, and varied NH$_3$ both above and below the 800-mbar level.  Once again, this produced the 2-12 ppm range of abundances at 440 mbar where CIRS sensitivity peaks.  However, CIRS prefers the deep $p>800$-mbar volume mixing ratio to vary from 350-400 ppm in the EZ, to 150-200 ppm in the NEB and 250-300 ppm in the SEB. This appears to be systematically higher than the values of \citeA{17li}.  Of the four tests above, test D provided the best fits to the CIRS spectra at all latitudes.    


Given the caveats above about the assumed temperatures in the MWR inversions, the new zonally-averaged CIRS reference model uses a latitudinally-uniform average of the MWR-derived deep NH$_3$ as a prior for $p>800$ mbar, but then allows NH$_3$ to vary both above and below this level to fit the spectra.  This is a minor update over that from \citeA{16fletcher_texes}, and still produces upper-tropospheric NH$_3$ abundances (at the 440-mbar level) of 2-12 ppm, depending on the latitude.  The equatorial enrichment is evident, alongside the asymmetric depletion of the NEB and SEB.  This new reference model is used to correct the radiometric calibration of TEXES in the main article.

\section{Stratospheric Results}
\label{app_stratos}

The TEXES equatorial inversions in Fig. \ref{zonal_texes} and \ref{zonal_temp} also provide information on the stratospheric temperatures and hydrocarbons.  Stratospheric temperatures near 1 mbar show an asymmetry between warm southern mid-latitudes and cool northern mid-latitudes, as expected from the bright but variable band of methane emission at 1248 cm$^{-1}$ between $10-20^\circ$S (i.e., above the SEB) in Fig. \ref{group5}.  The equator is cooler than the northern and southern bands, indicating the present phase of Jupiter's equatorial stratospheric oscillation in March 2017 \cite{91leovy} and consistent with IRTF mapping during the same period \cite{18melin}.  Stratospheric hydrocarbons ethane and acetylene show variability with longitude and a north-south asymmetry.  Both species show a local maximum at the equator that was hard to distinguish in previous IRTF observations \cite<e.g., see Fig. 18 of>{16fletcher_texes}, and may be indicative of local stratospheric circulation cells associated with the QQO.  A further maximum is present over the NTB ($21-28^\circ$N), previously seen in the acetylene distribution by \citeA{18melin} and suggestive of strong vertical mixing and/or heating at that latitude.  

\section{Windshear and Vorticity}
\label{windshear}

The presence of highly variable plumes and hot spots is responsible for the variability in the microwave brightness in the $5-10^\circ$N region in Fig. \ref{mwr_zonal}, and the thermal infrared variability in the same region in Fig. \ref{zonal_texes}.  Identifying a true zonally-averaged temperature or abundance in the presence of these NEBs features is rather challenging.  In our Supplemental Material (\url{https://doi.org/10.5281/zenodo.3706796}) we provide estimates, for each of the seven TEXES groups, of the thermal lapse rate, heat capacity, equilibrium para-H$_2$ fraction, atmospheric density, scale height, Brunt V\"{a}is\"{a}l\"{a} buoyancy frequency (a measure of the stratification), vertical windshear and thermal wind (i.e., integrating the thermal wind equation in altitude).  These 2D profiles (latitude, altitude) are available to those wishing to test their numerical simulations of the jovian tropics.

\begin{figure*}
\includegraphics[angle=0,width=1.0\textwidth]{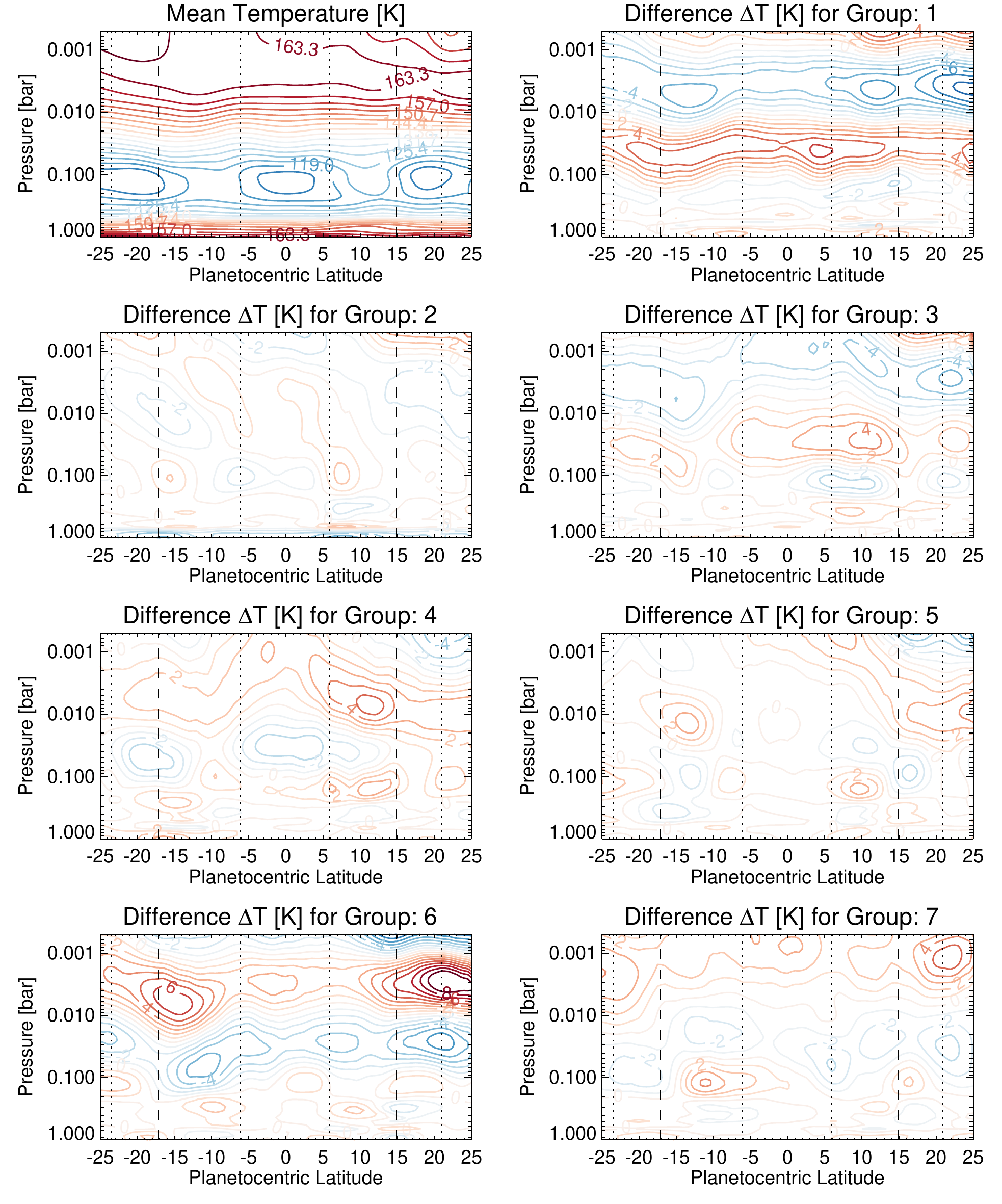}
\caption{Difference between the meridional temperatures in each of the seven TEXES groups (Table \ref{tab:data}) and the mean temperature (top left), showing the extent of the longitudinal variability. The contours go between $\pm10$ K in 1-K steps.}
\label{tempdiff}
\end{figure*}

\begin{figure*}
\includegraphics[angle=0,width=1.0\textwidth]{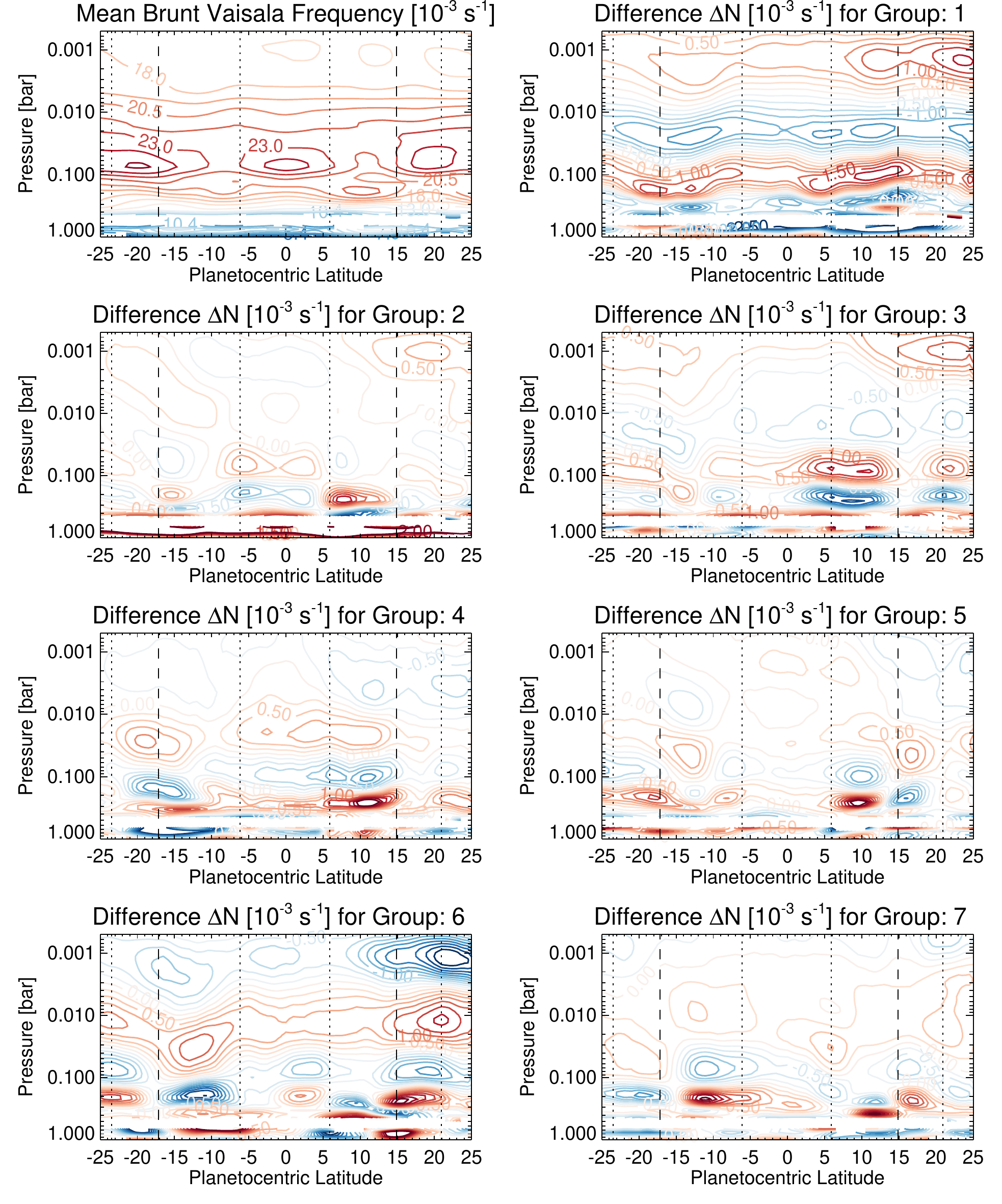}
\caption{Difference between the Brunt V\"{a}is\"{a}l\"{a} buoyancy frequency derived in each of the seven TEXES groups (Table \ref{tab:data}) and the mean buoyancy frequency (top left), showing the extent of the longitudinal variability. The contours go between $\pm2.5\times10^{-3}$ s$^{-1}$ in steps of $\pm0.25\times10^{-3}$ s$^{-1}$.}
\label{buoyancydiff}
\end{figure*}

The temperatures and winds derived from the seven TEXES groups show significant longitudinal variability, particularly in the $0-10^\circ$N range where the hot spots and plumes are located.  The Supplementary Information provides a comparison of the temperature inversions from the individual groups to the mean $T(p,\theta)$ (where $\theta$ is the latitude) in Fig. \ref{tempdiff}.  Temperature differences are smallest in the troposphere ($\Delta T<4$ K), but grow larger in the stratosphere ($\Delta T\sim10$ K in some extreme cases).  This is echoed in the buoyancy frequencies ($N$) shown in Fig. \ref{buoyancydiff}, where uncertainties are increased by taking the gradient of the vertical temperature profiles.  Regions where $N$ becomes imaginary (i.e., $N^2<0$) indicate statically unstable locations, whereas regions where $N^2>0$ are statically stable.  The TEXES inversions suggest that this occurs for pressures exceeding 400-500 mbar at all latitudes, which is near to the location of the radiative-convective boundary.

\begin{figure*}
\includegraphics[angle=0,width=1.0\textwidth]{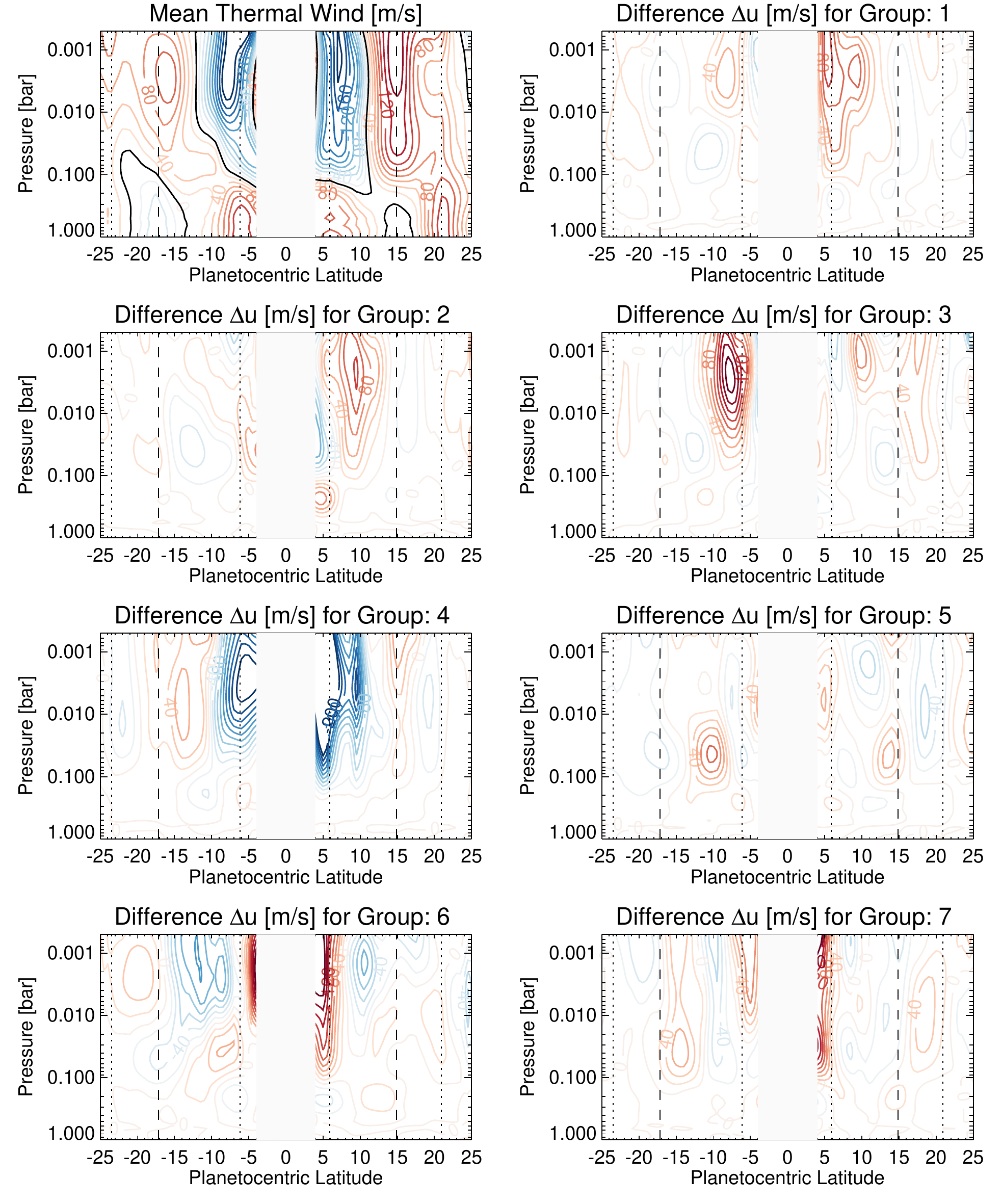}
\caption{Difference between the thermal wind derived in each of the seven TEXES groups (Table \ref{tab:data}) and the mean thermal wind (top left).  Uncertainties on the temperature gradients in Fig. \ref{tempdiff} cause large uncertainties in the integration of the thermal wind equation, meaning that the contours in this figure go between $\pm200$ m/s in steps of $20$ m/s.}
\label{winddiff}
\end{figure*}

Vertical windshears (derived from meridional temperature gradients) are integrated with altitude to indirectly estimate the thermal winds in Fig. \ref{winddiff}.  Given that the temperature profiles lack information content in the $20<p<80$ mbar range, there is a considerable uncertainty in the stratospheric winds \cite<see>[for a full discussion]{16fletcher_texes}.  Indeed, the derived winds are extremely sensitive to the meridional temperature differences shown in Fig. \ref{tempdiff}, meaning that estimates of winds can vary by up to 20-30 m/s in the troposphere and 100-200 m/s in the stratosphere, depending on the TEXES group chosen.  Longitudinal temperature contrasts can therefore have a significant effect on the integrated wind field, such that winds derived in this manner should be treated with considerable caution.

Finally, the curvature of the derived winds with latitude and altitude are used to estimate the meridional gradient of the quasi-geostrophic potential vorticity \cite<known as `effective beta,' or $\beta_e$,>{87andrews}, which is a powerful diagnostic tool for atmospheric dynamics \cite<please refer to Section 5 of >[for a detailed description of the equations used in this calculation]{16fletcher_texes}.  Large values of $\beta_e$ are associated with the prograde jets acting as barriers to meridional mixing, and reversals in the sign of $\beta_e$ are associated with the occurrence of atmospheric instabilities \cite{06read_jup}.  However, given the large differences in the thermal winds from longitude to longitude in Fig. \ref{winddiff}, the $\beta_e$ estimates provided in our Supporting Information are similarly variable and are only really meaningful as either a zonal average, or associated with an individual discrete feature.  These derived products for each TEXES group (winds and their vertical/horizontal derivatives) represent the limit of what can be diagnosed from ground-based facilities.

\section{Additional TEXES Hot Spots and Plumes}
\label{moretexes}

\begin{figure*}
\includegraphics[angle=0,width=1.0\textwidth]{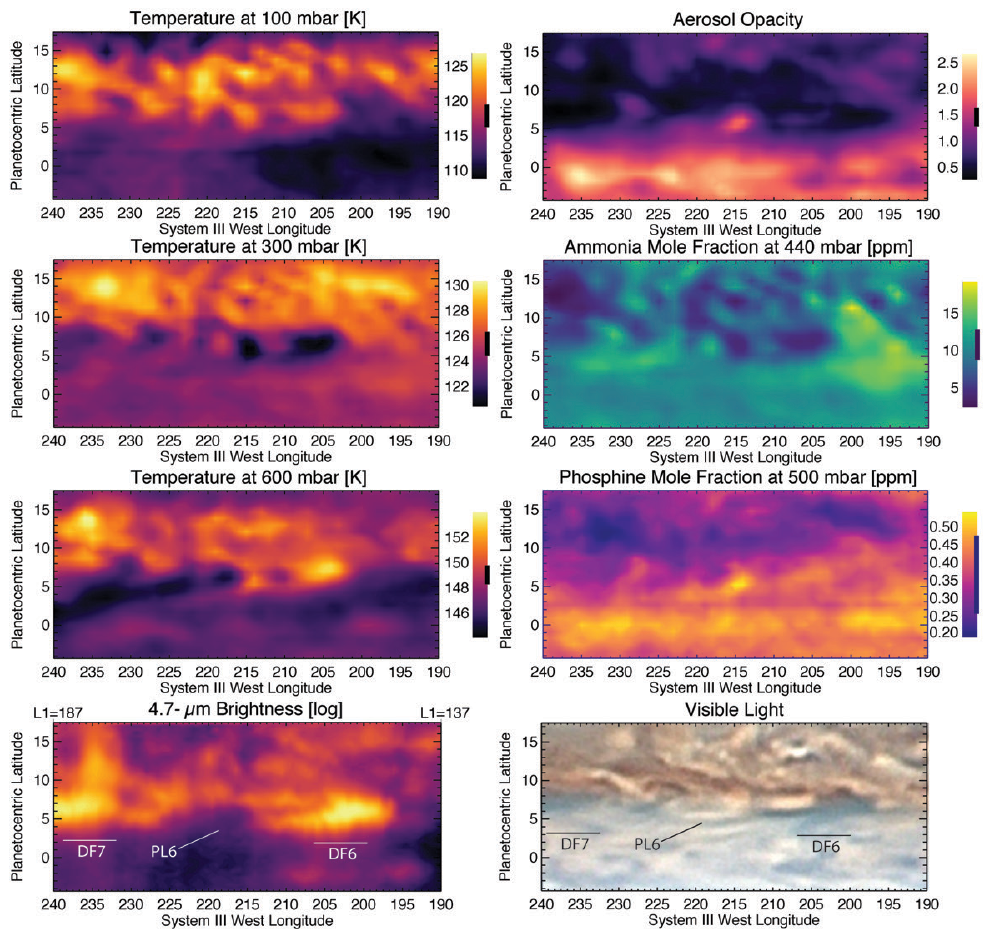}
\caption{Retrieved upper-tropospheric properties from TEXES group 1, 12 March 2017.  Temperatures at 100, 300, and 600 mbar are shown on the left, compared to the brightness at 4.7 $\mu$m (bottom left), which has been stretched logarithmically to reveal fainter features.  Retrieved aerosol opacity (cumulative optical depth to the 1-bar level) and ammonia and phosphine mole fractions are shown on the right, with a visible-light image from 20 hours earlier (T. Kumamori, 11 March 2017 at 15:15UT, adjusted in longitude to co-align the NEBs features).  We label the TWO dark formations (DF6 and 7) and a plume (PL6).  Black vertical lines on the legend for each figure show the formal retrieval uncertainty.  The System-I longitude range (L1) is added to the 4.7-$\mu$m map as a guide.}
\label{nebg1}
\end{figure*}

\begin{figure*}
\includegraphics[angle=0,width=1.0\textwidth]{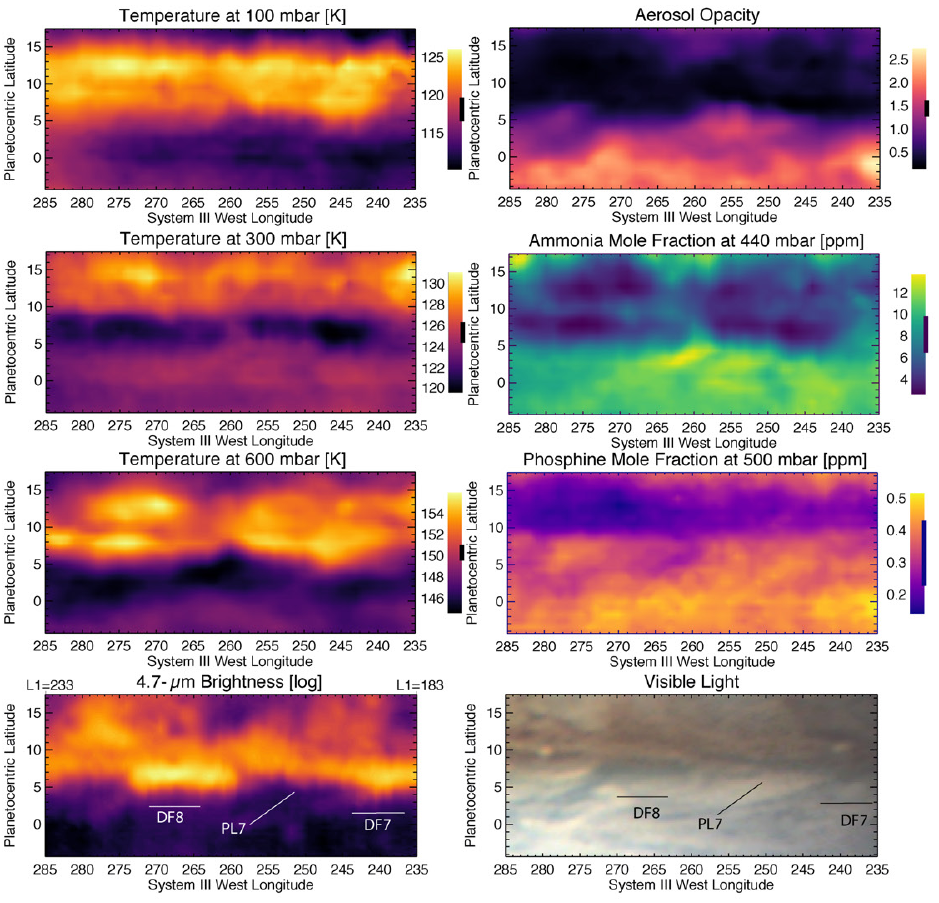}
\caption{Retrieved upper-tropospheric properties from TEXES group 2, 12 March 2017.  Temperatures at 100, 300, and 600 mbar are shown on the left, compared to the brightness at 4.7 $\mu$m (bottom left), which has been stretched logarithmically to reveal fainter features.  Retrieved aerosol opacity (cumulative optical depth to the 1-bar level) and ammonia and phosphine mole fractions are shown on the right, with a visible-light image from A. Soares taken 18 hours later (14 March 2017 04:42UT, adjusted in longitude to co-align the NEBs features).  We label the two dark formations (DF7 and 8) and a plume (PL7).  Black vertical lines on the legend for each figure show the formal retrieval uncertainty.  The System-I longitude range (L1) is added to the 4.7-$\mu$m map as a guide.}
\label{nebg2}
\end{figure*}

\begin{figure*}
\includegraphics[angle=0,width=1.0\textwidth]{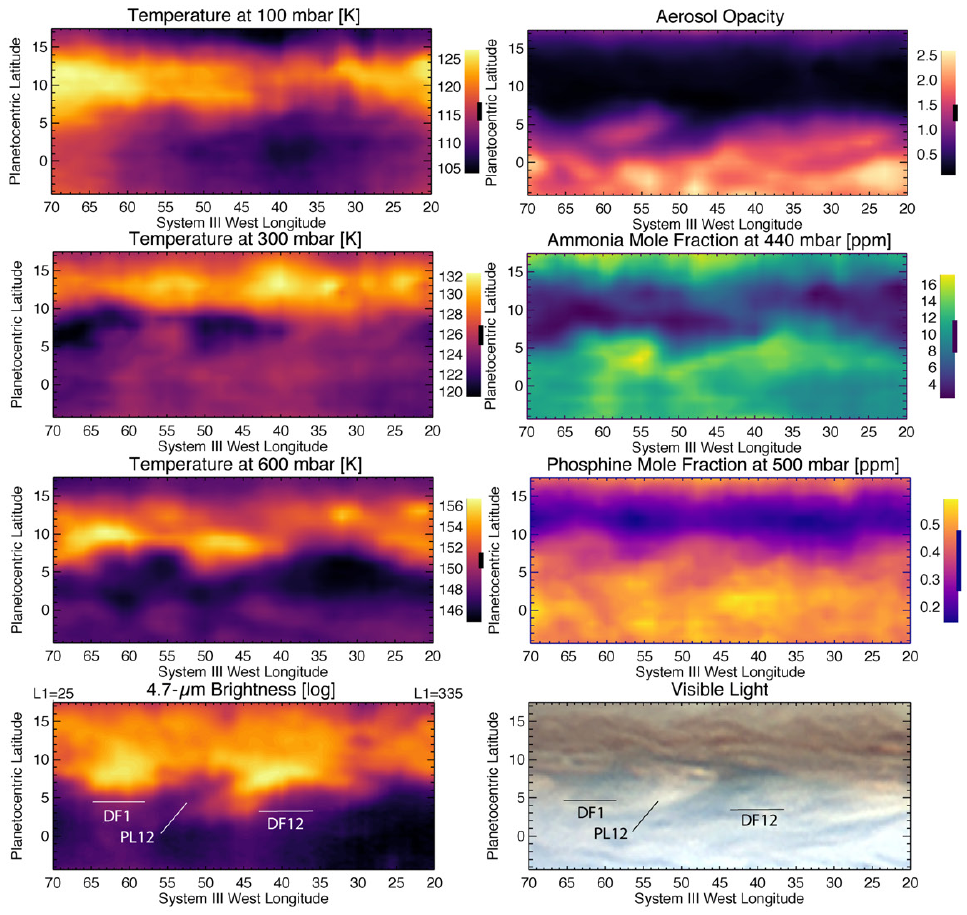}
\caption{Retrieved upper-tropospheric properties from TEXES group 5, 13 March 2017.  Temperatures at 100, 300, and 600 mbar are shown on the left, compared to the brightness at 4.7 $\mu$m (bottom left), which has been stretched logarithmically to reveal fainter features.  Retrieved aerosol opacity (cumulative optical depth to the 1-bar level) and ammonia and phosphine mole fractions are shown on the right, with a visible-light image from A. Garbelini Jr taken 30 hours earlier (12 March 2017 06:16UT, adjusted in longitude to co-align the NEBs features).  We label the two dark formations (DF12 and 1) and a plume (PL12).  Black vertical lines on the legend for each figure show the formal retrieval uncertainty.  The System-I longitude range (L1) is added to the 4.7-$\mu$m map as a guide.}
\label{nebg5}
\end{figure*}

In this appendix we provide retrieved maps of dark formations and plumes at locations that were not covered by Juno's perijoves in 2017 (groups 1, 2 and 5 in Table \ref{tab:data}, shown in Figs. \ref{nebg1}, \ref{nebg2} and \ref{nebg5}).  Groups 2 and 5 confirm the spatial variability in temperature/NH$_3$ between and within the individual hot spots that have been described in the main text.  The hot spot DF6 in group 1 (Fig. \ref{nebg1}, on 12 March 2017) shows some intriguing structure from east to west:  the 5-$\mu$m brightness extends from $197^\circ$W to $207^\circ$W, and is coincident with a local minimum in the cloud opacity at $\sim800$ mbar derived from the 8.6-$\mu$m observation.  This is also the location of the dark formation observed in visible light. However, the physical temperature maximum at 600 mbar, as well as the region of strongest NH$_3$ depletion, are further to the west, located near $203-211^\circ$W.  Given that all of the TEXES observations were taken within 70 minutes (Table \ref{tab:data}), this appears to be a real feature of the data.

It is possible that the eastern-most edge of the hot spot DF6 (Fig. \ref{nebg1}) is obscured by a notable region of enhanced NH$_3$ gas between $190-200^\circ$W.  Indeed, if the Galileo Probe had encountered a region like this, it would have potentially detected enriched ammonia compared to the surrounding NEB, and we would expect this region to appear dark in the microwave observations (which was not the case by July 2017).  This does suggest that the hot spot is spatially inhomogeneous along its $\sim10^\circ$ longitude extent, and that the composition depends on \textit{where} in the hot spot the observations are sounding.  DF7 is also bright at 5 $\mu$m, aerosol depleted, and shows subtle warming and NH$_3$ depletion compared to the surroundings.  But plume PL3, which is enriched in aerosols and dark at 5 $\mu$m, does not show particularly strong NH$_3$ enrichment.  Interestingly, PH$_3$ is enhanced in a localised feature near $215^\circ$W, $7^\circ$N, co-located with the highest aerosol opacity at 800 mbar - this is at the eastern edge of the dark plume PL6 at 5 $\mu$m, and might correspond to a source region for the aerosols that then spread westward.  However, apart from the enhancement in the EZ and general depletion over the NEB, PH$_3$ shows poor correlation with the discrete features observed at other wavelengths, and certainly does not show the depletion in the hot spots and enrichments in the plumes that we might have expected.

The hot spots and plumes in Fig. \ref{nebg1} are revealed to be complicated objects - spatially inhomogeneous in temperature and ammonia from east to west, rather than being uniformly depleted in ammonia (even though aerosol depletion/enrichment seems to span the whole hot spot/plume feature, respectively).  This again points to the need for MWR scans with broader longitudinal coverage to capture the full extent of a dark formation, and may explain why some MWR scans (which covered dark formations) did not necessarily show high brightness temperatures.

\acknowledgments
Fletcher is a Juno Participating Scientist supported by a Royal Society Research Fellowship and European Research Council Consolidator Grant (under the European Union's Horizon 2020 research and innovation programme, grant agreement No 723890) at the University of Leicester.  Orton is supported by funds from NASA distributed to the Jet Propulsion Laboratory, California Institute of Technology.

TEXES observations were obtained at the Gemini Observatory (program ID GN-2017A-Q-60), which is operated by the Association of Universities for Research in Astronomy, Inc., under a cooperative agreement with the NSF on behalf of the Gemini partnership: the National Science Foundation (United States), the National Research Council (Canada), CONICYT (Chile), Ministerio de Ciencia, Tecnología e Innovación Productiva (Argentina), and Ministério da Ciência, Tecnologia e Inovação (Brazil).  TEXES spectroscopic scan maps from March 2017 are available at \url{https://doi.org/10.5281/zenodo.3702328}.  Derived products from the TEXES observations, along with supporting images from amateur observers and IRTF/SpeX, are available on \url{https://doi.org/10.5281/zenodo.3706796}.

We are grateful to the worldwide network of amateur observers for their imaging during 2017, including those whose images were selected for use in this article:  I. Sharp, C. Foster, D. Peach, T. Olivetti, T. Kumamori, E. Martinez, A. Garbelini Jr., and S. Kidd.  We also thank M. Vedovato for the visible light map in Fig. \ref{map1165}.  SpeX observations were acquired at the IRTF during programs 2016B-077, 2016B-015, 2017A-034 and 2017A-035, and raw observations are available to download from the Juno-IRTF website\footnote{\url{https://junoirtf.space.swri.edu/}}.  Processed, mapped observations are available at \url{https://doi.org/10.5281/zenodo.3706796}.

Juno observations are available through the Planetary Data System Atmospheres Node\footnote{\url{https://pds-atmospheres.nmsu.edu/data_and_services/atmospheres_data/JUNO}}.  In addition,  JunoCam observations can be downloaded immediately from the Mission Juno web site\footnote{\url{https://www.missionjuno.swri.edu/junocam/}}.  Processed and cylindrically-mapped versions of the JunoCam, MWR, and JIRAM observations are available at \url{https://doi.org/10.5281/zenodo.3706793}.  The JIRAM project was funded by the Italian Space Agency (ASI), and we are grateful to all those who participated in the design of these observations.


%
%

\bibliography{references.bib}

%
%
%
%
%

\end{document}